\documentclass[prc,showpacs,superscriptaddress,twocolumn]{revtex4}

\usepackage{amsmath,bm}
\usepackage{graphicx}
\usepackage{epstopdf}
\usepackage{subfigure}
\usepackage{epsfig}
\usepackage{amsmath,amssymb,amsfonts}
\usepackage{enumitem}
\usepackage{color}
\usepackage[utf8]{inputenc}
\usepackage{verbatim}
\usepackage{array}
\usepackage{hyperref}
\graphicspath{{figures/}} % Directory in which figures are stored

\begin{document}

\renewcommand{\topfraction}{0.85}
\renewcommand{\textfraction}{0.1}
\renewcommand{\floatpagefraction}{0.75}
\def\be{\begin{eqnarray}}
\def\ee{\end{eqnarray}}
\newcommand{\mt}[1]{\textrm{\tiny #1}}
\def\pt{{p_\mt{T}}}
\def\ponet{{p_\mt{T1}}}
\def\ptwot{{p_\mt{T2}}}

\title{Unveiling the jet angular broadening with photon-tagged jets in high-energy nuclear collisions}

\author{Sa Wang}
\email{wangsa@ctgu.edu.cn}
\affiliation{College of Science, China Three Gorges University, Yichang 443002, China}
\affiliation{Center for Astronomy and Space Sciences and Institute of Modern Physics, China Three Gorges University, Yichang 443002, China}
\affiliation{Key Laboratory of Quark \& Lepton Physics (MOE) and Institute of Particle Physics, Central China Normal University, Wuhan 430079, China}

\author{Yao Li}
\affiliation{Key Laboratory of Quark \& Lepton Physics (MOE) and Institute of Particle Physics, Central China Normal University, Wuhan 430079, China}

\author{Jin-Wen Kang}
\affiliation{Key Laboratory of Quark \& Lepton Physics (MOE) and Institute of Particle Physics, Central China Normal University, Wuhan 430079, China}

\author{Ben-Wei Zhang}
\email{bwzhang@mail.ccnu.edu.cn}
\affiliation{Key Laboratory of Quark \& Lepton Physics (MOE) and Institute of Particle Physics, Central China Normal University, Wuhan 430079, China}

\date{\today}

%%%%%%%%%%%%%%%%%%%%%%%%%%%%%%%%%%%%%%%%%%%%%%%%%%%%%%%%%%%%%%%%%%%%%
\begin{abstract}
Medium modification of jet substructure in the hot and dense nuclear matter has garnered significant interest from the heavy-ion physics community in recent years. Measurements of inclusive jets show an angular narrowing in nucleus-nucleus collisions, while recent CMS results for photon-tagged jets ($\gamma$+jets) suggest evidence of broadening. In this study, we conduct a theoretical analysis of the angular structure of inclusive jets and $\gamma$+jets using a transport approach that accounts for jet energy loss and medium response in quark-gluon plasma. We examine the girth modification of $\gamma$+jets in $0-30\%$ PbPb collisions at $\sqrt{s_{NN}} = 5.02$ TeV, achieving good agreement with recent CMS measurements. We explore the relationship between selection bias and jet kinematics by varying the threshold for $x_{j\gamma} = p_T^{\rm jet}/p_T^{\gamma}$. Notably, we quantitatively demonstrate that $\gamma$+jets significantly reduce selection bias and can effectively select jets that have been sufficiently quenched in PbPb collisions, which is crucial for capturing the jet angular broadening. Additionally, we estimate the contributions of medium-induced gluon radiation and medium response to the broadening of the jet angular substructure. Finally, we analyze the modification patterns of jet $R_g$ and $\Delta R_{\rm axis}$ in PbPb collisions, which indicate slight broadening for $\gamma$+jets and noticeable narrowing for inclusive jets compared to pp collisions.
\end{abstract}

\maketitle

%%%%%%%%%%%%%%%%%%%%%%%%%%%%%%%%%%%%%%%%%%%%%%%%%%%%%%%%%%%%%%%%%%%%%
\section{Introduction}
\label{sec:introduction}

High-energy collisions of heavy nuclei at the Relativistic Heavy Ion Collider (RHIC) and the Large Hadron Collider (LHC) provide an experimental avenue to unravel the mysteries of quark-gluon plasma (QGP), a short-lived state of de-confined nuclear matter created at extremely high temperature and density. The jet quenching phenomenon, energy dissipation of an energetic parton when passing through hot and dense nuclear matter, is one of the most important signatures of the QGP formation~\cite{Gyulassy:2003mc, Gyulassy:1990ye, Qin:2015srf, Vitev:2008rz, Proceedings:2007ctk, Vitev:2009rd, Casalderrey-Solana:2014bpa, Gyulassy:1993hr,
Wang:2001ifa,Vitev:2008vk}. Investigations on jet quenching reveal the phase structure of the strongly-coupled nuclear matter and push our knowledge of the quantum chromodynamics (QCD) under extreme conditions~\cite{Connors:2017ptx, Andrews:2018jcm, Cao:2020wlm, Cunqueiro:2021wls, Apolinario:2022vzg, Sorensen:2023zkk, Zhao:2020wcd, Xie:2024xbn, Zhang:2021xib, Tang:2020ame, Shen:2020mgh, Chen:2024eaq, Chen:2024aom, Zhao:2022dac, Shou:2024uga}.

Jet substructures are valuable tools to gain insight into the details of the jet-medium interaction in QGP, such as the medium-induced gluon radiation~\cite{Baier:1996kr, Baier:1996sk}, medium response~\cite{KunnawalkamElayavalli:2017hxo, Pablos:2019ngg, Chen:2020tbl, Casalderrey-Solana:2020rsj, He:2018xjv, Ke:2020clc}, medium resolution length~\cite{Hulcher:2017cpt, Mehtar-Tani:2016aco, Caucal:2018dla} and the ``Moli\`{e}re elastic scattering''~\cite{DEramo:2012uzl, DEramo:2018eoy}. Recent reviews on this topic can be found in references~\cite{Cunqueiro:2021wls, CMS:2024krd, Arslandok:2023utm, Apolinario:2022vzg, Marzani:2019hun}. A significant focus of recent investigations has been on how the angular structure of jets is modified in nucleus-nucleus collisions—whether they narrow or broaden—an issue that has garnered much attention~\cite{Ringer:2019rfk, Rajagopal:2016uip, Chien:2016led, Mehtar-Tani:2016aco, Larkoski:2017bvj, Chang:2017gkt, Casalderrey-Solana:2016jvj, Kang:2023ycg, Kang:2023qxb, JETSCAPE:2023hqn, Milhano:2017nzm, Caucal:2019uvr, Casalderrey-Solana:2019ubu, Wang:2019xey, Li:2022tcr, Budhraja:2023rgo}. Measurements focused on the angular structure of inclusive jets indicate that jets become narrower in PbPb collisions at both the RHIC~\cite{STAR:2021kjt} and the LHC~\cite{ALargeIonColliderExperiment:2021mqf, ATLAS:2022vii, ALICE:2018dxf, ALICE:2023dwg, Ehlers:2022dfp, ATLAS:2023hso}, contrary to the theoretical expectation of intra-jet broadening~\cite{Ringer:2019rfk, Chien:2016led, Milhano:2017nzm}. In experiments, the medium modifications of the jet substructure are typically assessed by comparing two jet samples in PbPb and pp collisions that are selected within the same $p_T$ bins. The sufficiently quenched jets, which traverse a longer path length and undergo relatively larger energy loss in the QGP, may be less likely to pass the $p_T$ selection threshold in AA collisions, while some jets with insufficient quenching may still survive. This phenomenon is referred to as ``selection bias''~\cite{Baier:2001yt, Renk:2012ve, Wang:2021jgm, Cunqueiro:2021wls, Brewer:2021hmh, Zhang:2021sua}. Such biases could complicate the jet-by-jet comparison and obscure the connection between experimental measurements and intrinsic jet modifications~\cite{Wang:2021jgm, Brewer:2018dfs, Du:2020pmp, Brewer:2021hmh}.

The V+jet, jet tagged by the vector boson ($Z^0/W^{\pm}$ or $\gamma$), serves as a golden channel to explore the jet quenching phenomenon in high-energy heavy-ion collisions~\cite{Neufeld:2010fj, Wang:1996yh, Dai:2012am, Chen:2017zte, Wang:2013cia, Zhang:2018urd, Chang:2019sae, JETSCAPE:2024rma}. Since the vector boson does not interact strongly with the hot nuclear matter, it effectively gauges the initial momentum of the recoiling jet. Furthermore, the V+jet process is dominated by quark-jet production, reducing the potential influences from changes in the $q/g$ fraction during AA collisions~\cite{Wang:2020qwe}. Additionally constraining the $p_T$ of the vector boson was expected to minimize the impact of selection biases on jet measurements in AA collisions~\cite{Wang:2021jgm, Brewer:2021hmh, STAR:2023pal, STAR:2023ksv}. Therefore, V+jet may provide unique advantages to studying medium-induced jet broadening. Recently, the CMS collaboration reported the first measurement on the two angular structure observables of $\gamma$+jet, the jet girth ($g$)~\cite{Larkoski:2014pca} and the groomed jet radius ($R_g$)~\cite{Larkoski:2014wba}, in pp and PbPb collisions at $\sqrt{s}=5.02$ TeV~\cite{CMS:2024zjn}. The results show that the medium modification pattern of the angular structure significantly depends on the selection cut, $x_{j\gamma}=p_T^{\rm jet}/p_T^{\gamma}$, where $p_T^{\rm jet}$ and $p_T^{\rm \gamma}$ denote the transverse momentum of the jet and photon respectively. Notably, when setting $x_{j\gamma}>$ 0.4, there are hints of a broadening in jet angular structure at larger girth in PbPb collisions. This contrasts with the previously measured narrower girth distribution of inclusive jets reported by the ALICE collaboration~\cite{ALICE:2018dxf, ALargeIonColliderExperiment:2021mqf}. The influences of selection bias in these two types of jet measurements warrant careful consideration. Timely theoretical explanations and quantitative investigations are essential for addressing this issue.

This paper presents a theoretical study of the angular structure of $\gamma$+jets in high-energy nuclear collisions at the LHC. By employing a transport approach, we carry out the medium modification of $\gamma$+jet girth in $0-30\%$ PbPb collisions at $\sqrt{s_{NN}}=5.02$ TeV, which shows good agreement with recently reported CMS data. Using the Jet-by-Jet matching method, we explore the relationship between selection bias and kinematic requirements in realistic event selection. Through quantitative analysis, we demonstrate that $\gamma$+jets can significantly reduce selection bias and effectively select sufficiently quenched jets in PbPb collisions compared to inclusive jets. We also discuss the influence of medium-induced gluon radiation and medium response on the broadening of the jet angular substructure. For completeness, we will study the nuclear modification patterns of $R_g$ and $\Delta R_{\rm axis}$ distributions of both inclusive jets and $\gamma$+jets in PbPb collisions compared to the pp baseline.

The remainder of this paper is organized as follows. In Section \ref{sec:ppbaseline}, we will introduce the theoretical frameworks to study the angular structure of $\gamma$+jets in pp and PbPb collisions. In Section \ref{sec:res}, we will discuss the main results of this paper. At last, we will summarize this study in Section \ref{sec:sum}.
%%%%%%%%%%%%%%%%%%%%%%%%%%%%%%%%%%%%%%%%%%%%%%%%%%%%%%%%%%%%%%%%%%%%%
\section{Theoretical framework}
\label{sec:ppbaseline}

To investigate the angular structure of inclusive jets and $\gamma$+jets, we employ the PYTHIA8~\cite{Sjostrand:2014zea} with the Monash Tune~\cite{Skands:2014pea} to generate the pp events as a baseline for the calculations of nucleus-nucleus collisions. Furthermore, we utilize a transport approach, which considers both the radiative and collisional partonic energy loss, to simulate massive and massless jet evolution in QGP. This hybrid transport approach has been used in the studies of light- and heavy-flavor dijet~\cite{Dai:2018mhw, Li:2024uzk, Wang:2024yag, Li:2024pfi}, $Z^0/\gamma+$jet production in heavy-ion collisions~\cite{Wang:2020qwe, Wang:2023udp}. Since the medium-induced gluon radiation plays a critical role in jet energy loss~\cite{Gyulassy:1993hr, Wang:2001ifa}, we use the radiation spectrum within the higher-twist formalism~\cite{Guo:2000nz, Zhang:2003yn, Zhang:2003wk, Majumder:2009ge} to simulate the in-medium jet shower in the hot/dense QCD matter,

\begin{eqnarray}
\frac{dN}{ dxdk^{2}_{\perp}dt}=\frac{2\alpha_{s}C_sP(x)\hat{q}}{\pi k^{4}_{\perp}}\sin^2(\frac{t-t_i}{2\tau_f})(\frac{k^2_{\perp}}{k^2_{\perp}+x^2M^2})^4 \nonumber, \\
\label{eq:dndxk}
\end{eqnarray}
where $x$ and $k_\perp$ denote the energy fraction and transverse momentum carried by the radiated gluon. $\alpha_s$ is the strong coupling constant, $C_s$ the quadratic Casimir in color representation, $P(x)$ is the QCD splitting kernel~\cite{Deng:2009ncl} for $q\rightarrow q+g$ and $g\rightarrow g+g$ respectively($g\rightarrow q+\bar{q}$ process is negligible due to its low possibility~\cite{He:2011zx}),

\begin{eqnarray}
P_{q\rightarrow q+g}(x)=&\frac{(1-x)(1+(1-x)^2)}{x},\\
P_{g\rightarrow g+g}(x)=&\frac{2(1-x+x^2)^3}{x(1-x)}.
\end{eqnarray}
Moreover, $\tau_f=2Ex(1-x)/(k^2_\perp+x^2M^2)$ is the gluon formation time considering the Landau-Pomeranchuk-Migdal (LPM) effect~\cite{Wang:1994fx,Zakharov:1996fv}. $\hat{q}=q_0(T/T_0)^3$ denotes the jet transport parameter~\cite{Chen:2010te, Rapp:2018qla, Cao:2018ews}, where $T_0$ is the medium temperature at the center of QGP at $\tau=0.6$ fm. In this work, we use the extracted value of $q_0=1.2$ GeV$^2$/fm determined with a $\chi^2$ fitting to the identified hadron production in PbPb collisions at LHC~\cite{Ma:2018swx}. To consider the fluctuation of medium-induced gluon radiation, we assume that the number of the radiated gluon during a time step ($\Delta t=0.1$ fm) obeys the Poisson distribution $f(n)=\lambda^{n}e^{-\lambda}/{n!}$, where the parameter $\lambda$ denotes the mean number of the radiation calculated by integrating Eq.~(\ref{eq:dndxk}). Once the radiation number $n$ is determined, the corresponding energy-momentum can be further sampled by Eq.~(\ref{eq:dndxk}) one by one. Due to the LPM effect, a radiated gluon can only interact independently with the medium after a formation time $\tau_f$ and then further lose energy (including medium-induced gluon radiation). Note that Eq.~(\ref{eq:dndxk}) represents the net radiative contribution at the high-energy limit as derived from the complete twist-4 calculations~\cite{Chen:2010te, Guo:2000nz, Zhang:2003yn, Zhang:2003wk}. We impose a lower energy cutoff for the radiated gluons at the Debye screening mass $\mu_D$ to avoid the infrared behavior of Eq.~(1) in the limit $x\rightarrow 0$, where $\mu_{D}^2=4\pi\alpha_s(1+N_f/6)T^2$ with $N_f=3$.

In addition, it is also essential to consider the partonic energy loss from the elastic scattering. While the inelastic jet energy loss is described using higher-twist formalisms, for completeness, the elastic energy loss is estimated via pQCD calculations within the Hard-Thermal-Loop (HTL) approximation at leading logarithmic accuracy~\cite{Braaten:1991we, Neufeld:2010xi},
\begin{eqnarray}
\frac{dE}{dL}=-\frac{C_s\alpha_s g_s^2T^2}{2}(1+\frac{N_f}{6}){\rm ln}{\frac{q_{\rm max}}{q_{\rm min}}}, \nonumber
\label{eq:HTL}
\end{eqnarray}
 where $L$ represents the transport path along the parton's momentum direction. We set physically reasonable upper and lower limits for the 3-momentum transfer $q$ as $q_{\rm max}\sim\sqrt{ET}$ and $q_{\rm min}\sim m_D$. The collisional energy loss of a parton can be calculated by integrating Eq.~(\ref{eq:HTL}) during each time step $\Delta t$ according to the parton energy and the temperature of the medium cell. This treatment is an adequate approximation since the medium-induced gluon radiation is the dominant energy loss mechanism for light partons. The initial spatial production vertex of jets in nucleus-nucleus collisions is determined based on the MC-Glauber model~\cite{Miller:2007ri}. In the simulation of a jet traversing the expanding fireball, we utilize the CLVisc hydrodynamic model~\cite{Pang:2016igs} to generate the temperature and velocity of each medium cell. When the local temperature reaches $T_c=0.165$~GeV, jet partons fragment into hadrons with the Colorless Hadronization prescription, which the JETSCAPE collaboration developed~\cite{Putschke:2019yrg} based on the Lund string model~\cite{Andersson:1983jt, Sjostrand:1984ic}.

When studying jet substructures in high-energy nuclear collisions, it is important to consider the medium response effect. Energy transferred from high-$p_T$ jets can excite quasi-particles within the QGP medium~\cite{KunnawalkamElayavalli:2017hxo, Pablos:2019ngg, Chen:2020tbl, Casalderrey-Solana:2020rsj, He:2018xjv, Ke:2020clc}. In this work, we utilize an approach based on the Cooper-Frye formula with perturbations~\cite{Cooper:1974mv, Casalderrey-Solana:2016jvj} to take the medium response effect into account.

\begin{eqnarray}
E\frac{d\Delta N}{d^3p}&=&\frac{m_T}{32\pi T^5}{\rm cosh(\Delta y)exp}[-\frac{m_T}{T}{\rm cosh(\Delta y)]} \nonumber\\
&&\times \{p_T\Delta P_T{\rm cos(\Delta \phi)}+\frac{1}{3}m_T\Delta M_T{\rm  cosh(\Delta y)}\} \nonumber.\\
\label{eq:resp}
\end{eqnarray}

Here, $\Delta y$ and $\Delta \phi$ represent the rapidity and azimuthal angle of the emitted thermal particles relative to the jet axis, respectively. Meanwhile, $m_T$ and $p_T$ denote their transverse mass and transverse momentum. The quantities $\Delta P_T$ and $\Delta M_T = \Delta E / \cosh y_j$ signify the transverse momentum and transverse mass transferred from the jet to the medium, where $\Delta E$ is the energy lost by the jets, and $T$ indicates the hadronization temperature of the emitted particles. Once the values for $\Delta P_T$ and $\Delta E$ of the jet during in-medium propagation are determined, one can sample the transverse momentum, rapidity, and azimuthal angle of the emitted particles individually based on Eq. (\ref{eq:resp}). All emitted partons from the excited medium are assumed to hadronize into protons and pions ($\pi^0,\pi^{\pm}$) with a ratio of $5\%$ for protons and $95\%$ for pions. This branching ratio was approximately extracted in the Hybrid model~\cite{Casalderrey-Solana:2016jvj, Casalderrey-Solana:2020rsj} by fitting the particle spectra from PbPb collisions at $\sqrt{s_{NN}}=5.02$ TeV measured by the ALICE collaboration~\cite{ALICE:2013mez}. In that work, the hadronization temperature $T$ was parameterized using the following species- and momentum-dependent empirical expressions.

\begin{eqnarray}
&&\hspace{-0.3in}
T_{\pi}(\pt)=\begin{cases} 0.19~{\rm GeV}, &  \pt<0.7~{\rm GeV}\\
0.21 \, \left(\frac{\pt}{\rm{GeV}}\right)^{0.28} \, \rm{GeV}, &  \pt>0.7~{\rm GeV} \end{cases}\\
&&\hspace{-0.3in}T_p(\pt)=\begin{cases}
0.15~{\rm GeV}, &  \pt<0.07~{\rm GeV}\\
0.33 \, \left(\frac{\pt}{\rm{GeV}}\right)^{0.3} \rm{GeV}, &  0.07\sim 1.9~{\rm GeV}\\
0.4~{\rm GeV}, &  \pt>1.9~{\rm GeV} \end{cases}
\end{eqnarray}

%%%%%%%%%%%%%%%%%%%%%%%%%%%%%%%%%%%%%%%%%%%%%%%%%%%%%%%%%%%%%%%%%%%%%
\section{Results and discussions}
\label{sec:res}

Recently, the medium modifications of $\gamma$+jet girth distribution in PbPb collisions $\sqrt{s_{NN}}=5.02$ TeV are measured by the CMS collaboration~\cite{CMS:2024zjn}.
The jet girth belongs to a generalized class of observables known as jet angularities~\cite{Larkoski:2014pca, KunnawalkamElayavalli:2017hxo}, defined as,

\begin{eqnarray}
g=\frac{1}{p_T^{\rm jet}}\sum_{i \in \rm jet} p_T^i \Delta R_{i,\rm jet} ,
\label{eq:girth}
\end{eqnarray}
where the index $i$ sums over all jet constituents, and the $\Delta R_{i,\rm jet}$ represents the angular distance between each particle and the jet axis in the $\eta-\phi$ plane. The jet axis is determined by summing the momenta of all constituent particles of the anti-$k_T$ jet using the ``E-scheme'' clustering algorithm~\cite{Cacciari:2008gp}. The girth value quantifies the $p_T$ distribution of particles within a jet, weighted by their angular distance from the jet axis. This observable is expected to be sensitive to modifications in the jet angular structure that occur during heavy-ion collisions~\cite{ALICE:2018dxf, CMS:2024zjn, Yan:2020zrz, Li:2024pfi}. The jet-medium interactions, such as the parton scattering, medium-induced gluon radiation, and medium response, may increase the values of jet girth in heavy-ion collisions compared to those in pp.

\begin{figure}[!t]
\begin{center}
%\hspace*{-0.1in}
\includegraphics[width=3.0in,angle=0]{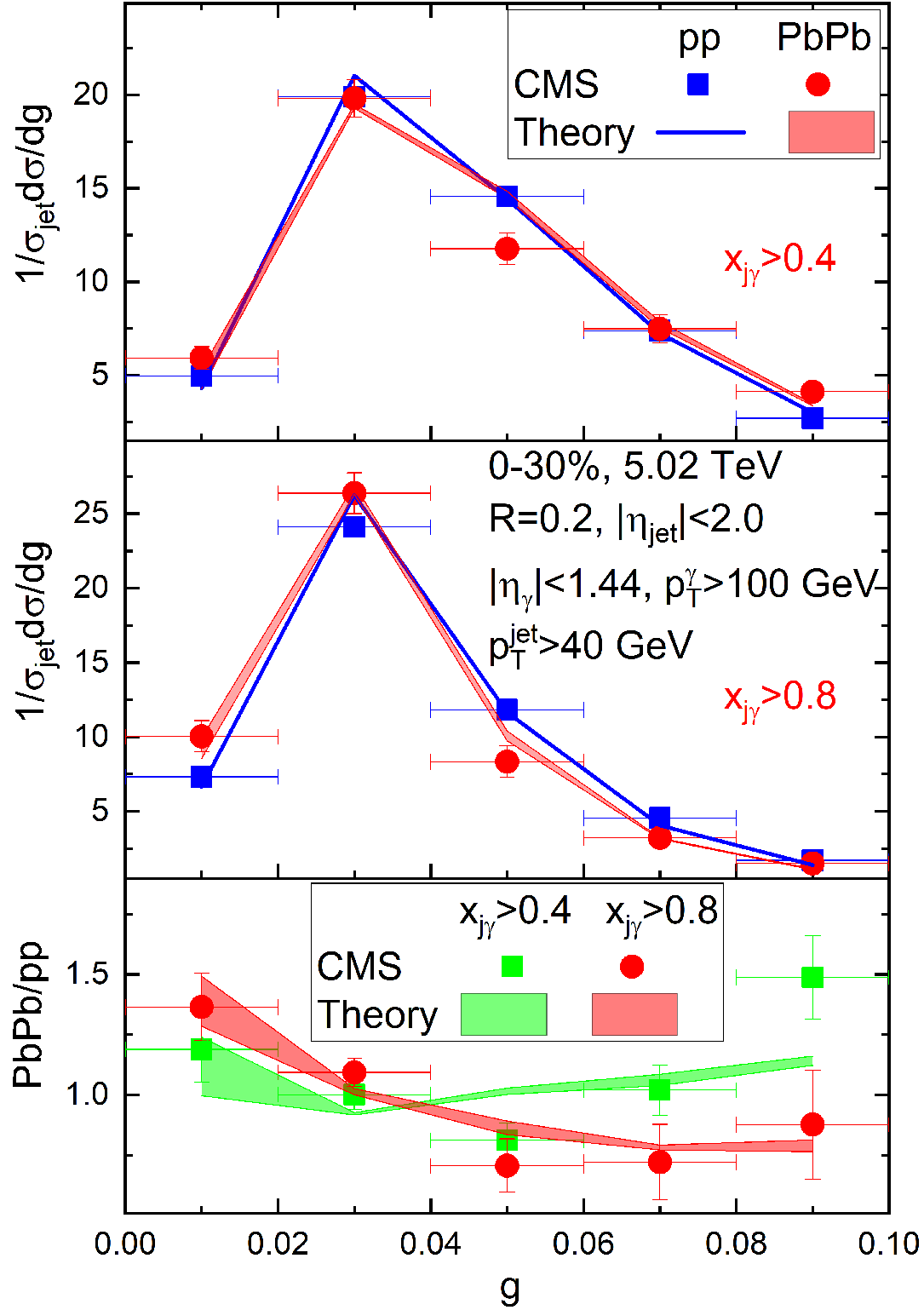}
\caption{(Color online) Normalized girth distributions of $\gamma$+jets in pp $0-30\%$ PbPb collisions at $\sqrt{s_{NN}}=5.02$ TeV calculated with $x_{j\gamma}>$ 0.4 (upper panel) and $x_{j\gamma}>$ 0.8 (middle panel) as compared to the recent CMS data~\cite{CMS:2024zjn}. The ratios (PbPb/pp) of girth distribution are also shown in the lower panel.}
\label{fig:ppAA-cms2}
\end{center}
\end{figure}

As shown in Fig. \ref{fig:ppAA-cms2}, we present the theoretical results of the $\gamma$+jet girth distribution for $x_{j\gamma} > 0.4$ (upper panel) and $x_{j\gamma} > 0.8$ (middle panel) in both pp and $0-30\%$ PbPb collisions at $\sqrt{s_{NN}} = 5.02$ TeV, compared with CMS data. The lower panels show the ratio PbPb/pp distributions. All selected jets are constructed with the anti-$k_T$ algorithm and are required to have a transverse momentum of $p_T^{\rm jet} > 40$ GeV, with pseudorapidity $|\eta_{\rm jet}| < 2$ and jet radius $R=0.2$, while the photon must have $p_T^{\gamma} > 100$ GeV and $|\eta_{\gamma}| < 1.44$. In addition, the selected photon must meet the isolation requirement, which states that the sum of the transverse momentum of all particles within a distance $R = 0.4$ around the photon must be less than 5 GeV. Furthermore, the jet and photon should be nearly ``back-to-back'', requiring $\Delta \phi_{\gamma,\rm jet} > 2\pi/3$. The medium evolution was simulated over a statistically significant ensemble of $10^7$ jet events, thereby guaranteeing that the relative statistical error in our theoretical calculations is negligible, quantified at less than $1\%$. This high precision provides a solid foundation for the subsequent analysis and discussion. Our theoretical calculations provide a satisfactory description of the CMS girth distribution data in both pp and PbPb collisions for the cuts $x_{j\gamma} > 0.4$ and $x_{j\gamma} > 0.8$~\cite{CMS:2024zjn}. Notably, the ratio of girth distributions between PbPb and pp collisions (PbPb/pp) differs significantly for the two $x_{j\gamma}$ cuts. For $x_{j\gamma} > 0.4$, the modification of girth is modest, with an enhancement observed between $0.08 < g < 0.1$. Since girth quantifies the angular-weighted transverse momentum distribution of jets, an increase in PbPb/pp at a larger girth indicates that jets are broader in PbPb collisions compared to pp collisions. In contrast, for $x_{j\gamma} > 0.8$, we observe an enhancement at $g > 0.02$ and a clear suppression at $g > 0.04$, suggesting that jets are narrower compared to the condition $x_{j\gamma} > 0.4$.

The error band of PbPb calculations shown in Fig.~\ref{fig:ppAA-cms2} comes from the statistical error of numerical computation and the uncertainty in the transport parameter $q_0=1.2\pm 0.2$ GeV$^2/$fm. Our calculations show that the statistical error is negligible (less than $1\%$) due to the large sample of over $10^7$ simulated jet events. The uncertainty associated with $q_0$ has a more pronounced impact in the smaller girth region, but it does not affect the broadening/narrowing patterns of the jet girth observed for $x_{j\gamma}>0.4$ and $x_{j\gamma}>0.8$. Thus, the difference in the nuclear modification patterns between the two $x_{j\gamma}$ cuts is robust and not due to theoretical uncertainties. For simplicity, in the subsequent discussion, we will only focus on the theoretical calculation with the central value $q_0=1.2$ GeV$^2/$fm. Results shown in Fig. \ref{fig:ppAA-cms2} prompts the question: how does the $x_{j\gamma}$ selection cut affect the patterns of jet angular structure modifications in nucleus-nucleus collisions? One possible explanation is that the lower $x_{j\gamma}$ cut allows for the inclusion of more significantly quenched jets traversing QGP, which reduces the selection bias effect~\cite{CMS:2024zjn}. To explore this hypothesis, we conduct a Jet-by-Jet matching procedure to investigate the relationship between selection bias and kinematic cuts in heavy-ion collisions.

\begin{figure}[!t]
\begin{center}
%\hspace*{-0.1in}
\includegraphics[width=3.0in,angle=0]{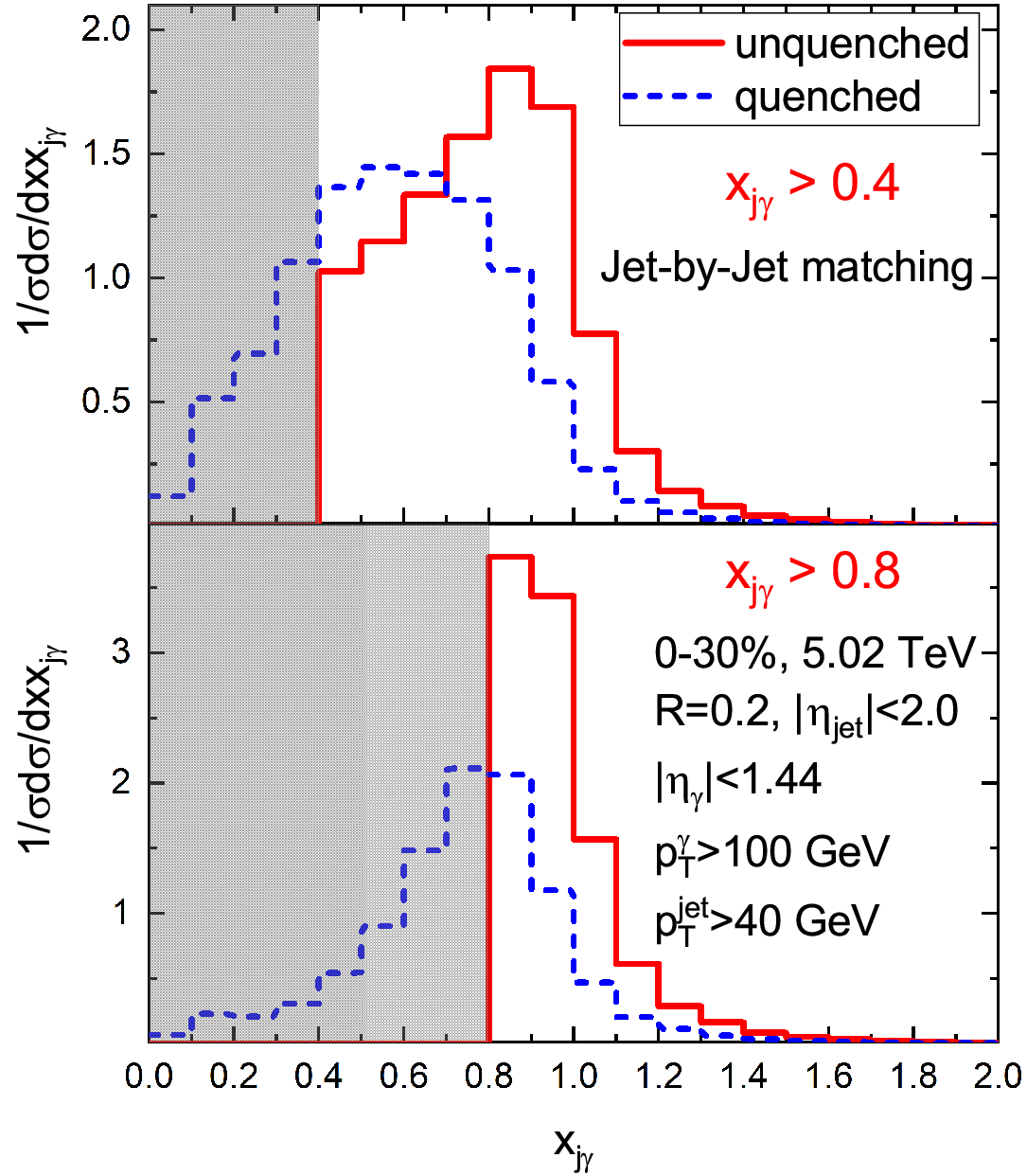}
\vspace*{0in}
\caption{(Color online) Normalized $x_{j\gamma}$ distribution of $\gamma$+jets before and after quenching, denoted as unquenched and quenched, in $0-30\%$ PbPb collisions at $\sqrt{s_{NN}}=5.02$ TeV when using different $x_{j\gamma}$ cuts by using the Jet-by-Jet matching procedure, $x_{j\gamma} > 0.4$ (upper panel) and $x_{j\gamma} > 0.8$ (lower panel).}
\label{fig:biasxj}
\end{center}
\end{figure}

In experimental measurements within the field of high-energy heavy-ion collisions, the conventional approach to studying the nuclear modification effects on jet substructure involves applying identical dynamical constraints during jet reconstruction and selection for both proton-proton and nucleus-nucleus collisions. By comparing the selected jet samples from these two collision systems, one can calculate the nuclear modification effects for the specific jet observables, such as jet girth. However, since the jets in these two samples do not correspond to the same individual jet before and after quenching, the derived nuclear modification effects represent an overall statistical average across a large ensemble of jets. Therefore, it is natural to inquire about how jet-medium interactions affect the substructure of individual jets before and after quenching in nucleus-nucleus collisions. While experimentally tracking the evolution of a single jet through different stages remains highly challenging, such processes can indeed be modeled and analyzed via a jet-by-jet matching in the Monte Carlo simulations as performed in the studies of Refs.~\cite{Brewer:2021hmh, Kang:2023ycg}. The details are illustrated as follows:

\textit{Jet-by-Jet matching}: The events selected with suitable experimental kinematic cuts, such as $p_T>$ 40 GeV and $|\eta_{jet}|<$ 2.0, in pp collisions are used as the input of jet evolution in PbPb. We can reconstruct the jets in pp and PbPb collisions by the event particles before and after the in-medium evolutions. We calculate the angular distance $\Delta R<R$ between the axes of each jet pair in pp and PbPb events. The jet pair with the smallest $\Delta R$ is then regarded as the matched one before and after quenching. To consider the jets, with $p_T>$ 40 GeV initially, dropping down to the cut due to jet energy loss, we use a lower cut $p_T>$ 10 GeV to select the possible candidates in PbPb collisions. This Jet-by-Jet (JBJ) matching method effectively eliminates selection bias, enabling precise tracking of the medium evolution of individual jets in the Monte Carlo simulations. It offers a novel perspective to study jet-medium interactions in high-energy heavy-ion collisions and provides critical insights for interpreting recent experimental measurements of jet substructure modifications.

 As shown in Fig.~\ref{fig:biasxj}, with the help of the Jet-by-Jet matching, we compare the $x_{j\gamma}$ distribution of $\gamma$+jets before and after quenching in $0-30\%$ PbPb collisions when using different $x_{j\gamma}$ cuts, $x_{j\gamma} > 0.4$ (upper panel) and $x_{j\gamma} > 0.8$ (lower panel). The solid line represents the initially selected (unquenched) jets above $x_{j\gamma}$ cut, while the dashed line denotes the corresponding jet sample after the in-medium evolution (quenched). Due to energy loss in QGP, the $x_{j\gamma}$ distributions of quenched jets shift toward lower $x_{j\gamma}$ values compared to the unquenched ones. The shaded region represents the jets that have suffered sufficient quenching effect in nucleus-nucleus collisions and have dropped down to the $x_{j\gamma}$ cut. The contribution of the shaded region was usually discarded in the realistic experimental measurements in PbPb collisions, even though they initially have $p_T > 40$ GeV. Therefore, the fraction of the shaded region quantitatively reflects the influence of selection bias, namely the amount of jets with sufficient quenching effect but are rejected by the selection requirement in the measurement in PbPb collisions. We find that the fraction of the shaded region is $23.9\%$ for $x_{j\gamma} > 0.4$ and $58.4\%$ for $x_{j\gamma} > 0.8$. It means that when applying the criterion $x_{j\gamma} > 0.8$, less than half of the jets survive the event selection in PbPb collisions, thereby excluding a significant number of quenched jets. In contrast, using $x_{j\gamma} > 0.4$ includes more jets that experience substantial quenching in PbPb collisions, ultimately resulting in a broader modification of the $\gamma$+jet girth compared to pp, as observed in the CMS measurement~\cite{CMS:2024zjn}.

\begin{figure}[t]
\begin{center}
%\hspace*{-0.1in}
\includegraphics[width=3.0in,angle=0]{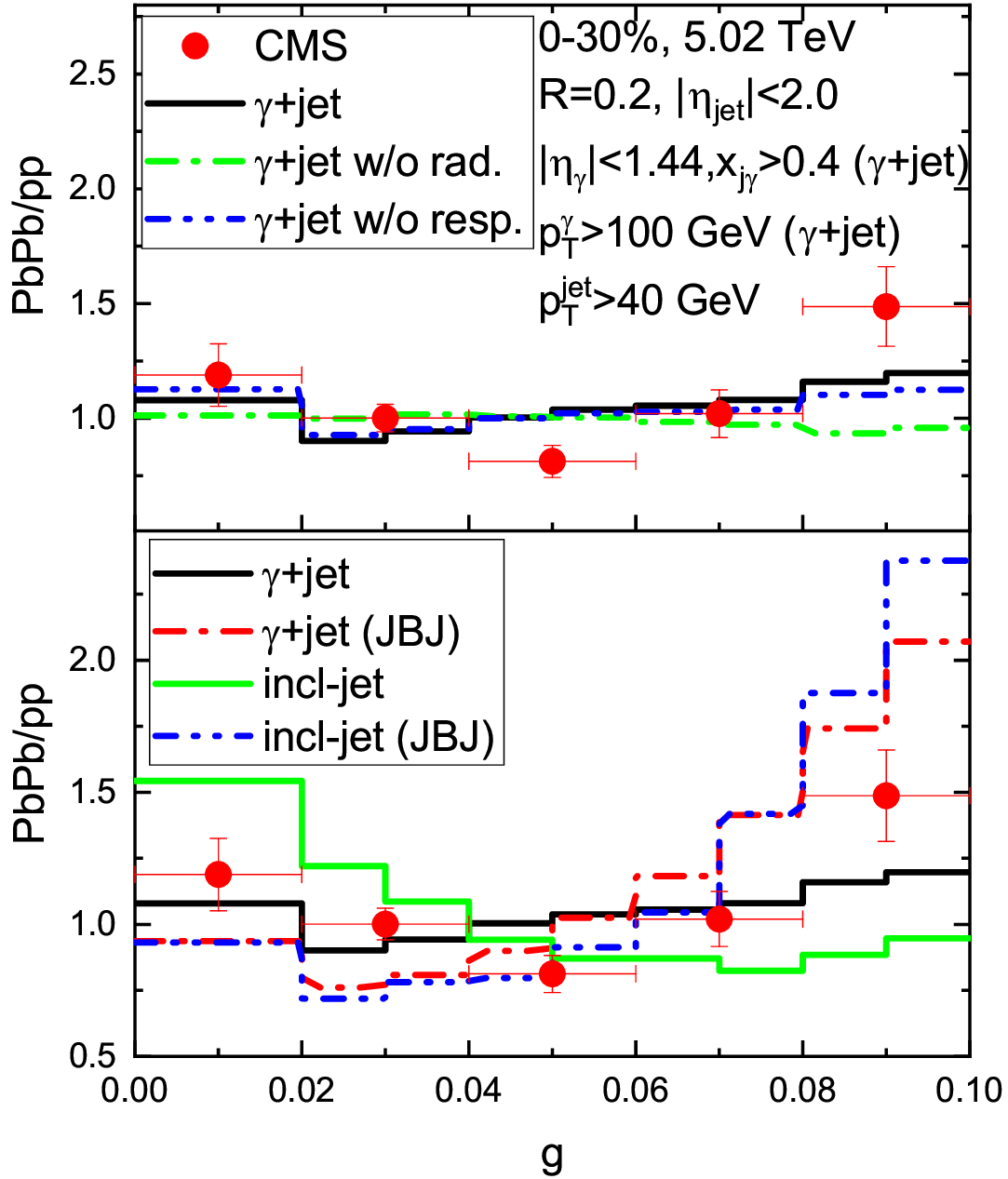}
\vspace*{0in}
\caption{(Color online) Medium modification of $\gamma$+jet girth in $0-30\%$ PbPb collisions relative to pp at $\sqrt{s_{NN}}=5.02$ TeV as compared to the CMS data~\cite{CMS:2024zjn}. In the upper panel, the results without considering the medium-induced gluon radiation and medium response are also presented. In the lower panel, we also compare the girth modification of inclusive jet and $\gamma$+jet with the same $p_T$ cut, as well as the case using the Jet-by-Jet matching (JBJ).}
\label{fig:gijetg}
\end{center}
\end{figure}

Furthermore, measurements of inclusive jets in PbPb collisions indicate a narrowing effect~\cite{ALargeIonColliderExperiment:2021mqf, ATLAS:2022vii, ALICE:2018dxf, ALICE:2023dwg, Ehlers:2022dfp, ATLAS:2023hso}, while observations from $\gamma$+jets suggest hints of broadening~\cite{CMS:2024zjn}. It will be of great significance to explore the differing patterns of substructure modification between these two jet samples within the same collision system. In Fig. \ref{fig:gijetg}, we present the girth modification of $\gamma$+jets and inclusive jets in $0-30\%$ PbPb collisions at $\sqrt{s_{NN}} = 5.02$ TeV. Aside from the tagging requirements for $\gamma$+jets—specifically, $p_T^{\gamma} > 100$ GeV, $x_{j\gamma} > 0.4$, and $|\eta_{\gamma}| < 1.44$—all selected jets must have $p_T > 40$ GeV and $|\eta_{\text{jet}}| < 2$. In the upper panel, we first discuss the influence of medium-induced gluon radiation and medium response on $\gamma$+jets. Compared to the results without gluon radiation, our findings indicate that medium-induced gluon radiation plays a significant role in the jet angular broadening in nucleus-nucleus collisions. Additionally, we also find that medium response slightly enhances the modification in the region of larger girth. In the lower panel, we compare the girth modifications of $\gamma$+jets and inclusive jets. We observe that inclusive jets show suppression at $g > 0.05$, which indicates a narrowing modification consistent with previous ALICE measurements in PbPb collisions at $\sqrt{s_{NN}} = 2.76$ TeV~\cite{ALICE:2018dxf}. This behavior differs from that of $\gamma$+jets. To address this discrepancy, we calculate the girth modification of the jet sample using the Jet-by-Jet matching method, which represents the jet modification in the hot QCD matter without selection bias. Both $\gamma$+jets and inclusive jets show consistent and noticeable broadening at larger girth compared to the initial jets. In other words, jets intrinsically become broader due to jet-medium interactions in PbPb collisions for both inclusive jets and $\gamma$+jets. We will demonstrate that selection bias plays different roles in $\gamma$+jets and inclusive jets, ultimately leading to the distinct modification patterns observed in experimental measurements.

\begin{figure}[t]
\begin{center}
%\hspace*{-0.1in}
\includegraphics[width=3.0in,angle=0]{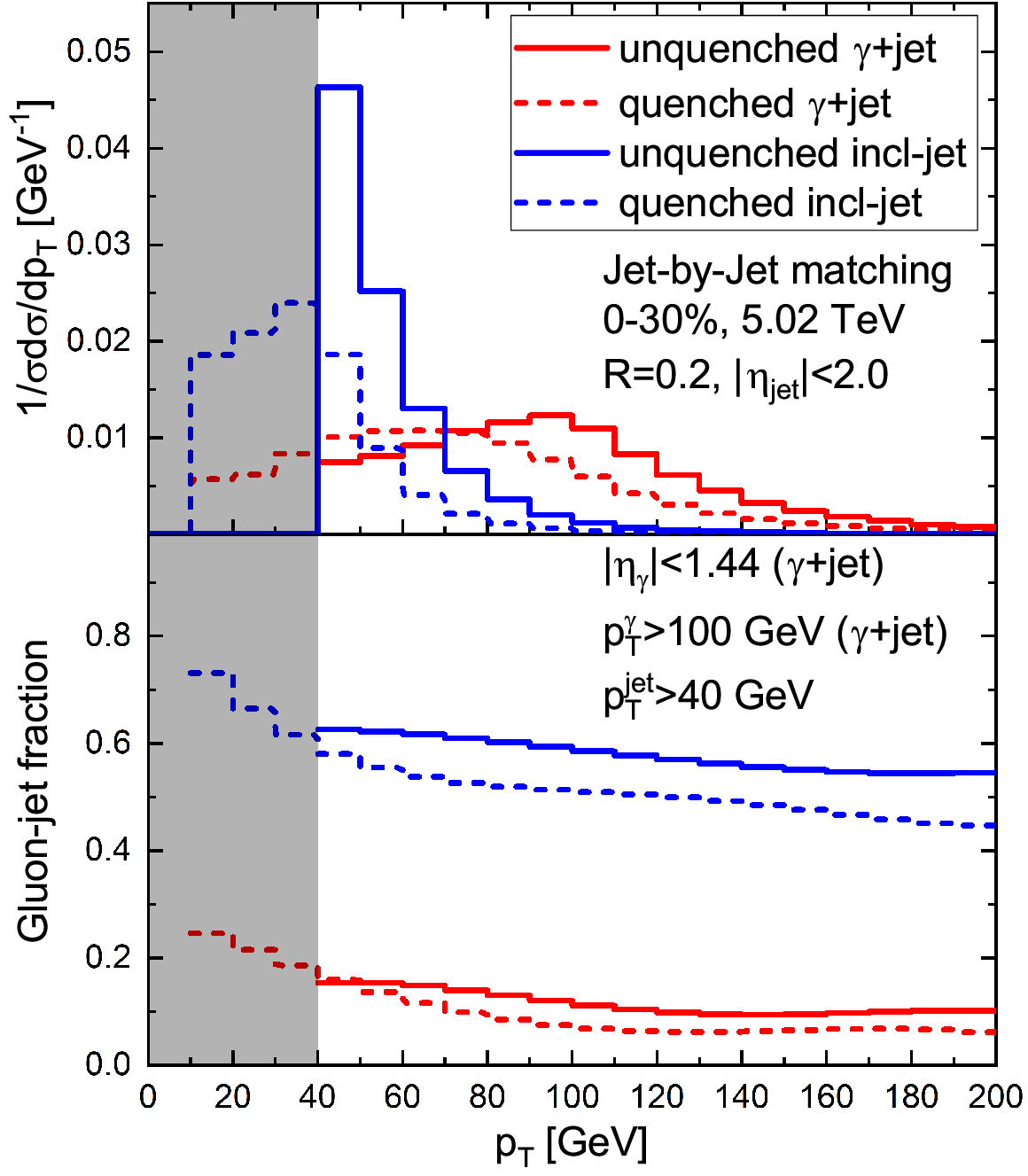}
\vspace*{0in}
\caption{(Color online) Normalized $p_T$ distribution of unquenched and quenched jets for $\gamma$+jet and inclusive jet in $0-30\%$ PbPb collisions at $\sqrt{s_{NN}}=5.02$ TeV by using the Jet-by-Jet matching procedure. The gluon-jet fractions in inclusive jet and $\gamma+$jet samples are also estimated in the lower panel.}
\label{fig:biasgjet}
\end{center}
\end{figure}

In Fig. \ref{fig:biasgjet}, we also compare the $p_T$ spectra of the jets before and after quenching in PbPb collisions for both $\gamma$+jets and inclusive jets, obtained using the JBJ matching. By applying the same jet $p_T$ cut (specifically, $ p_T > 40$ GeV), we observe that the shape of the $\gamma$+jet $p_T$ spectrum before quenching significantly differs from that of the inclusive jets. The $\gamma$+jet spectrum increases gradually with $p_T$, peaking near the photon $p_T$. In contrast, the inclusive jet distribution is mainly concentrated in the range of $[40, 80]$ GeV and decreases rapidly with increasing $p_T$. Due to in-medium energy loss, we find that the $p_T$ distributions of both $\gamma$+jets and inclusive jets in PbPb collisions shift towards a lower energy region compared to their pp counterparts. The shaded region indicates the jets that do not meet the \( p_T > 40 \) GeV requirement due to energy loss in PbPb. Specifically, the fraction of the shaded region is $20.3\%$ for $\gamma$+jets and $63.4\%$ for inclusive jets. In other words, approximately $80\%$ of $\gamma$+jets survive the selection criteria in PbPb, whereas less than $40\%$ of inclusive jets do. It indicates that using jets associated with direct photons can significantly reduce selection bias, as can be understood from two perspectives. First, inclusive jets are mostly distributed near the selection cut (40 GeV), while jets associated with photons cover a broader $p_T$ range with a peak near the $p_T$ of the triggered photon (100 GeV). As a result, the unique $p_T$ spectrum of the $\gamma$+jets provides a much lower probability of jets falling below the cut-off after quenching compared to the inclusive jets. Second, the inclusive jet sample contains a significant fraction of gluon jets, whereas the $\gamma+$jet process is dominated by quark jets. As shown in the lower panel of Fig.~\ref{fig:biasgjet}, the initial fraction of gluon jets in the inclusive jet sample is approximately $60\%$. This fraction decreases after quenching because the gluon's larger color charge leads to greater energy loss compared to quarks. In contrast, the gluon jet fraction in the $\gamma+$jet sample is significantly lower (around $10-20\%$ initially) and remains at about $10\%$ after quenching. Consequently, inclusive jets may lose more energy on average, resulting in a lower survival rate for passing the selection criteria in PbPb collisions.

\begin{figure}[t]
\begin{center}
%\hspace*{-0.2in}
\includegraphics[width=3.1in,angle=0]{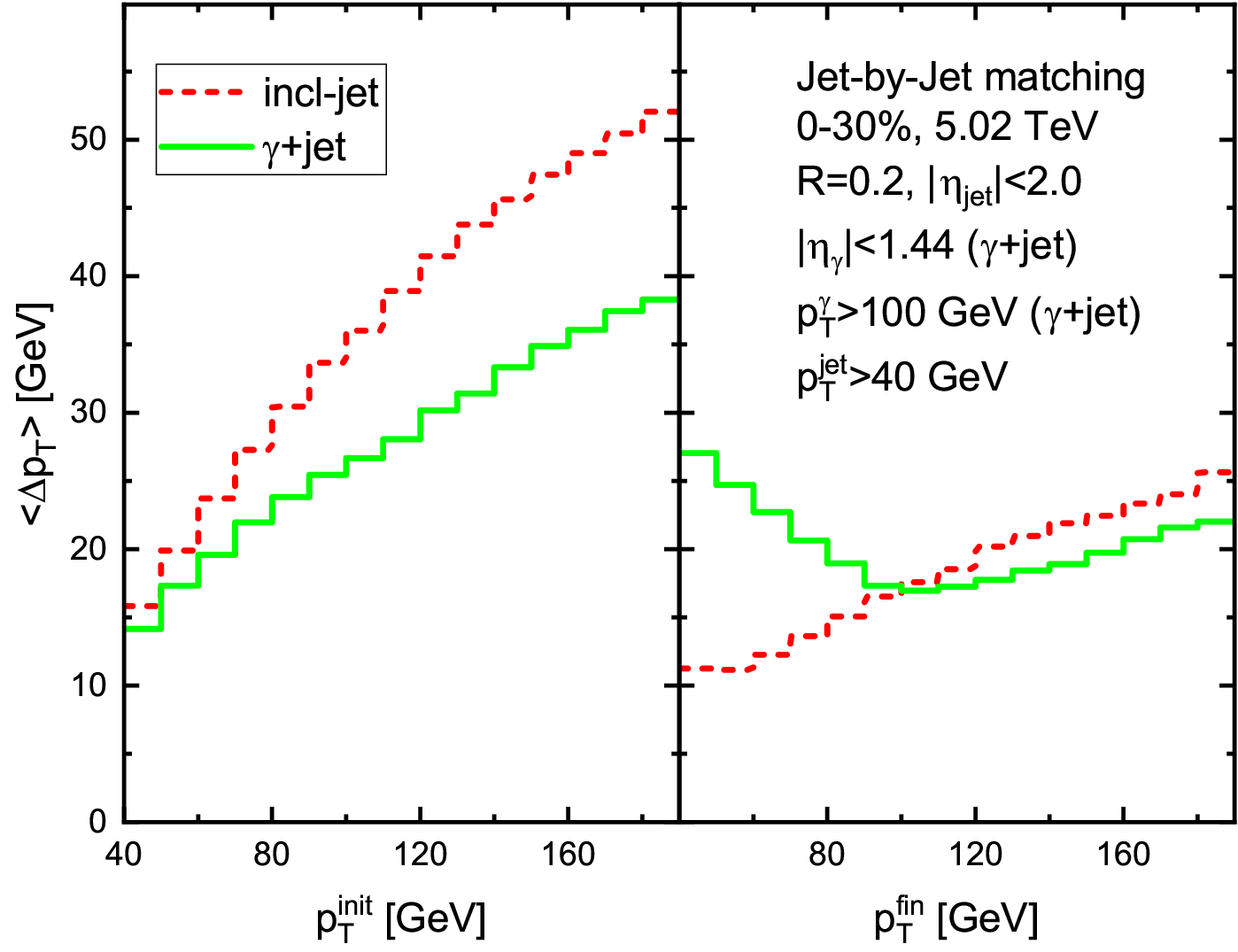}
\vspace*{-0in}
\caption{(Color online) Event averaged transverse momentum loss $\langle \Delta p_T \rangle=\langle p_T^{\rm init}-p_T^{\rm fin}\rangle_{\rm evt}$ of $\gamma$+jets (solid line) and inclusive jets (dash line) as a function of initial (left panel) and final (right panel) jet $p_T$ in $0-30\%$ PbPb collisions at $\sqrt{s_{NN}}=5.02$ TeV by using the Jet-by-Jet matching procedure.}
\label{fig:dpt}
\end{center}
\end{figure}

In addition to the higher survival probability of $\gamma$+jets compared to inclusive jets, we are also interested in understanding the differences in the quenching strength experienced by the surviving samples for these two types of jets within QGP. Since quenching strength is not a specifically defined physical quantity, we assume that jet energy loss can effectively quantify the intensity of the interaction between the jet and the medium. Therefore, we estimate the event-averaged $p_T$ loss, denoted as $\langle \Delta p_T \rangle = \langle p_T^{\rm init} - p_T^{\rm fin} \rangle_{\text{evt}}$, for both $\gamma$+jets and inclusive jets as a function of final jet $ p_T $ ($ p_T^{\rm fin}$) in $ 0-30\% $ PbPb collisions at $ \sqrt{s_{NN}} = 5.02 $ TeV, using Jet-by-Jet matching as illustrated in Fig. \ref{fig:dpt}. In the left panel, we plot the $ p_T $ loss of $\gamma$+jets and inclusive jets as a function of their initial transverse momentum. It can be observed that $ \langle \Delta p_T \rangle $ for both $\gamma$+jets and inclusive jets increases with $ p_T^{\rm init} $, with the inclusive jets showing larger magnitudes. It aligns with expectations, as inclusive jets contain a larger fraction of gluon jets, while $\gamma$+jets are predominantly comprised of quark jets. In the right panel of Fig. \ref{fig:dpt}, we also plot the $ \langle \Delta p_T \rangle $ for $\gamma$+jets and inclusive jets as a function of $ p_T^{\rm fin} $. Surprisingly, we find that $ \langle \Delta p_T \rangle $ for $\gamma$+jets is significantly larger than that for inclusive jets in the range of $ 40 < p_T^{\rm fin} < 80 $ GeV. It implies that selected $\gamma$+jets with $ p_T > 40$ GeV in PbPb collisions have statistically experienced stronger quenching than inclusive jets. Compared to inclusive jets, the specific initial $p_T$ distribution of $\gamma$+jets, which includes a substantial number of jets with much higher $p_T$ than the selection threshold, allows for the possibility that jets experiencing significant quenching can still survive in the selection in PbPb collisions. Therefore, we have quantitatively demonstrated that $\gamma$+jets provide unique and significant advantages in reducing selection bias and effectively collecting sufficiently quenched jets in PbPb collisions compared to inclusive jets.

\begin{figure}[t]
\begin{center}
%\hspace*{-0.1in}
\includegraphics[width=3.0in,angle=0]{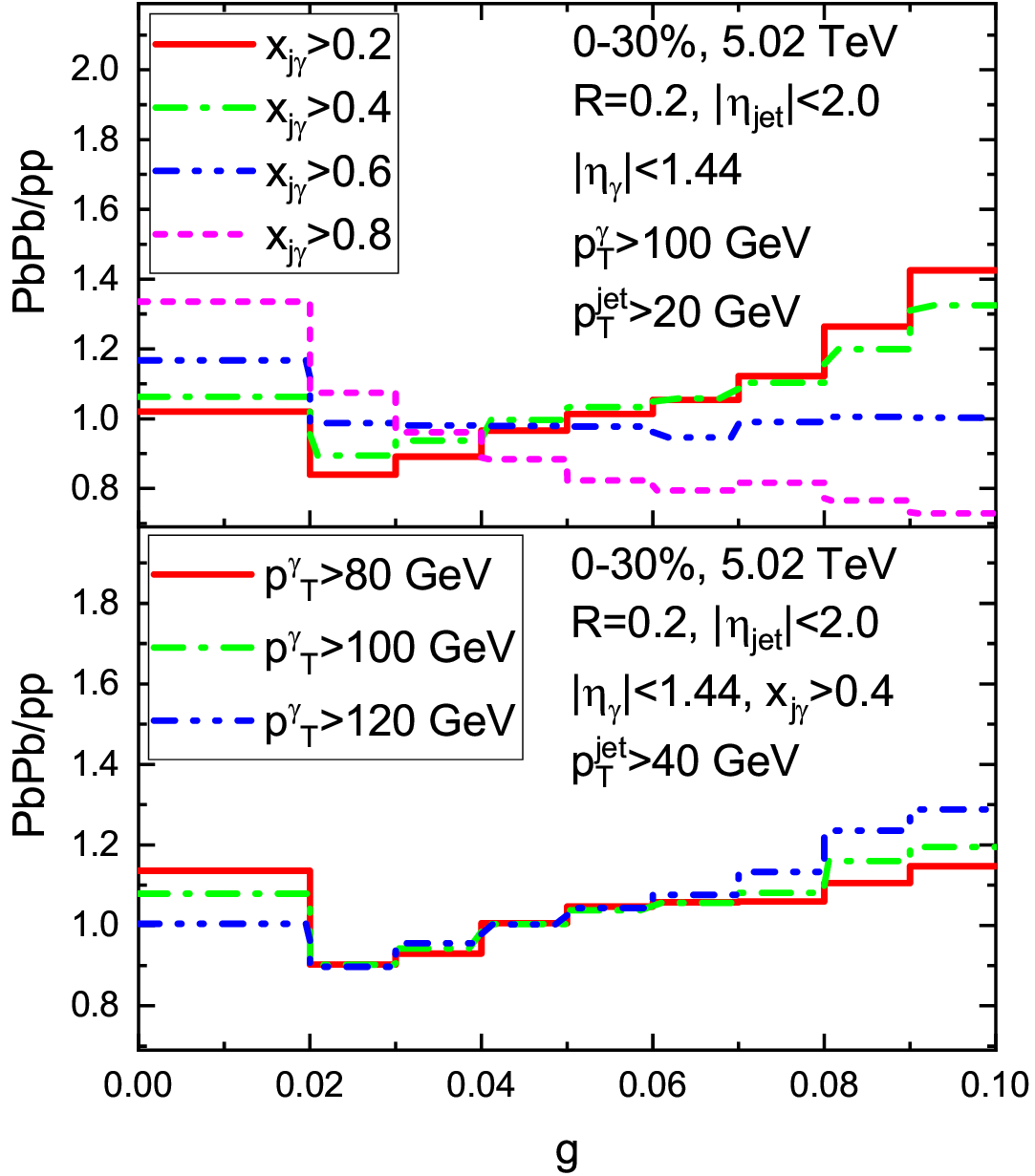}
\vspace*{0in}
\caption{(Color online) Medium modification of jet girth of $\gamma$+jets in $0-30\%$ PbPb collisions at $\sqrt{s_{NN}}=5.02$ TeV relative to pp for different $x_{j\gamma}$ (0.2, 0.4, 0.6, 0.8) and $p_{T}^{\gamma}$ (80 GeV, 100 GeV, 120 GeV) cuts.}
\label{fig:gxgpt}
\end{center}
\end{figure}

The discussions above highlight the connection between selection bias and the choice of the threshold $x_{j\gamma}=p_T^{\rm jet}/p_T^{\gamma}$. To offer theoretical guidance for searching for more pronounced angular broadening effects of $\gamma$+jets in experimental settings, we have conducted calculations that explore the sensitivities of jet girth modifications to the $x_{j\gamma}$ and $p_T^{\gamma}$ cuts, as illustrated in Fig.~\ref{fig:gxgpt}. In the upper panel, we calculate the girth modification of $\gamma$+jet in PbPb collisions with varying $x_{j\gamma}$ cuts. It is evident that when $x_{j\gamma}>0.2$, more pronounced angular broadening is observed because more quenched jets are accepted in the selection process in PbPb. As the $x_{j\gamma}$ cut increases from 0.2 to 0.8, the increasing influence of selection bias leads to a transition from enhancement at $g>0.06$ to suppression, namely the modification pattern of the jet changes from broadening to narrowing, consistent with the trend observed in CMS data~\cite{CMS:2024zjn}. Note that we have applied a lower $p_T$ cut of 20 GeV to reconstruct the jets, ensuring that the variation of $x_{j\gamma}$ down to 0.2 remains accessible. Additionally, the lower panel presents calculations of $\gamma$+jet girth modification in PbPb collisions with different $p_T^{\gamma}$ cuts. We find that applying a higher $p_T$ cut for the photon can also help mitigate the influence of selection bias and result in more pronounced girth modifications of $\gamma$+jets in PbPb collisions.

\begin{figure}[t]
\begin{center}
%\hspace*{-0.1in}
\includegraphics[width=3.0in,angle=0]{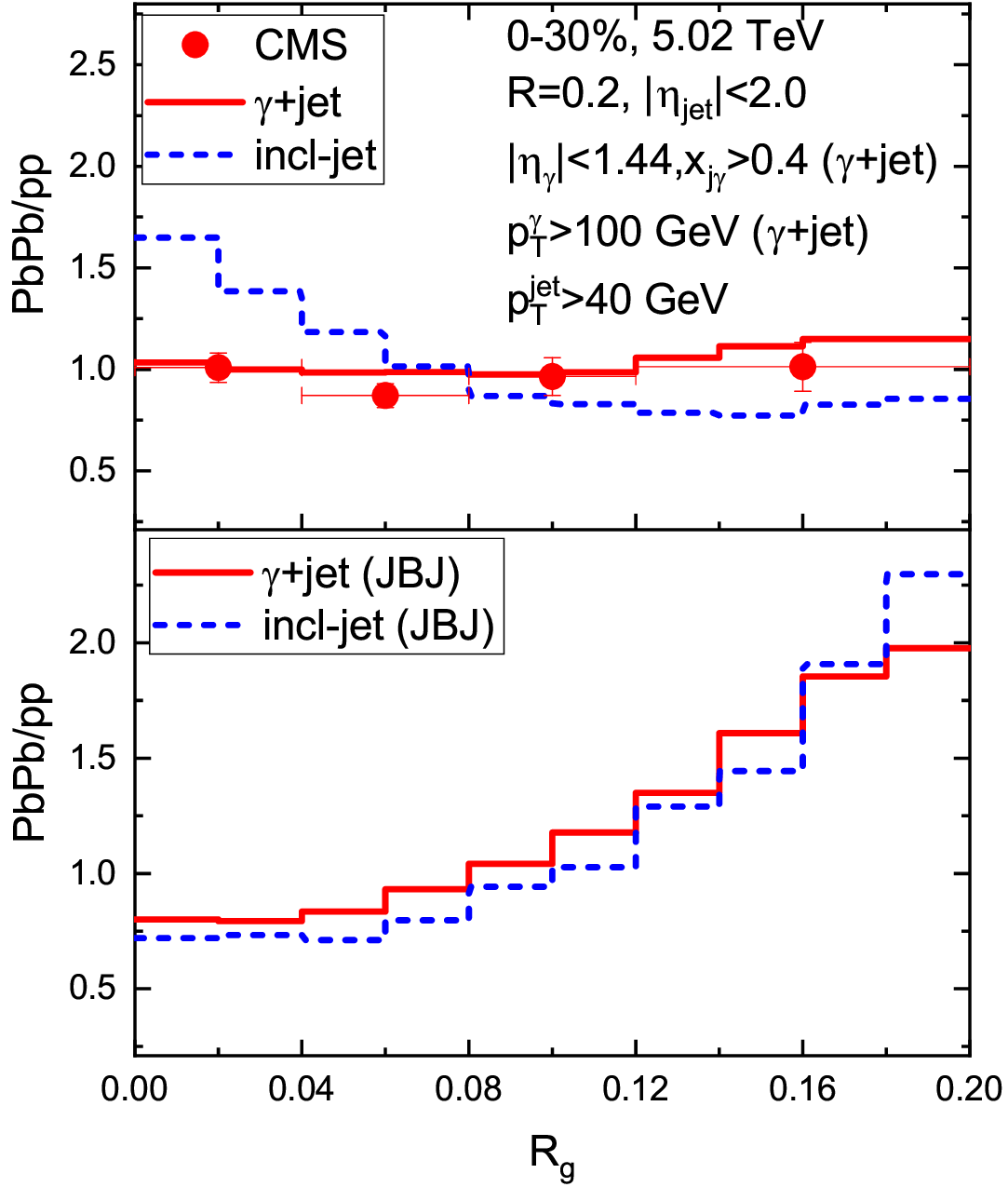}
\vspace*{0in}
\caption{(Color online) Medium modification of the groomed radius distribution of $\gamma$+jet and inclusive jet in $0-30\%$ PbPb collisions relative to pp at $\sqrt{s_{NN}}=5.02$ TeV. The results are compared with the measurements of $\gamma$+jet by the CMS~\cite{CMS:2024zjn}. We also plot the results using the Jet-by-Jet matching (JBJ) as a comparison in the lower panel.}
\label{fig:gijetRg}
\end{center}
\end{figure}

To provide a comprehensive analysis, we will also examine other jet substructure observables and highlight their different modification patterns for $\gamma$+jets and inclusive jets, similar to our findings regarding jet girth. In Fig.~\ref{fig:gijetRg}, we compare the medium modification of the groomed radius ($R_g$) distribution for $\gamma$+jets and inclusive jets in $0-30\%$ PbPb collisions relative to pp at $\sqrt{s_{NN}} = 5.02$ TeV. The PbPb/pp refers to the ratio of normalized $R_g$ distributions in PbPb to pp collisions. Here, $R_g$ represents the opening angle between the two hardest subjets (denoted as $1$ and $2$) that satisfy the Soft Drop condition~\cite{Larkoski:2014wba},

 \begin{eqnarray}
&&\frac{p_{T2}}{p_{T1}+p_{T2}}>z_{\rm cut}(\frac{R_{g}}{R})^\beta, \\
&&R_g=\sqrt{(\eta_1-\eta_2)^2+(\phi_1-\phi_2)^2}.
\end{eqnarray}
where $p_{T1}$ and $p_{T2}$ represent the transverse momentum of the leading and subleading subjets, respectively, while $R$ is the jet size. $\eta$ and $\phi$ are the pseudorapidity and azimuthal angle of the two subjets. We set $z_{\text{cut}} = 0.2$ and $\beta = 0$ in line with the CMS measurement~\cite{CMS:2024zjn}. The theoretical results for the modification of the groomed radius of $\gamma$+jets show a moderate enhancement at $R_g > 0.15$, consistent with CMS data, which may indicate jet angular broadening. In contrast, we observe that the medium modification pattern for inclusive jets shows a notable narrowing, diverging from the behavior of $\gamma$+jets. Additionally, by utilizing Jet-by-Jet matching as shown in the lower panel of Fig.~\ref{fig:gijetRg}, we can see an enhancement of the PbPb/pp ratio at $R_g > 0.1$ for both inclusive jets and $\gamma$+jets. This observation is analogous to what we discussed earlier regarding jet girth in Fig.~\ref{fig:gijetg}. Both inclusive jets and $\gamma$+jets become broader when they pass through QGP. However, selection bias reduces the broadening effect seen in the $R_g$ distribution of inclusive jets and $\gamma$+jets in nucleus-nucleus collisions. Despite this suppression, $\gamma$+jets exhibit a slight broadening effect, which is less influenced by selection bias compared to inclusive jets.

\begin{figure}[t]
\begin{center}
%\hspace*{-0.1in}
\includegraphics[width=3.0in,angle=0]{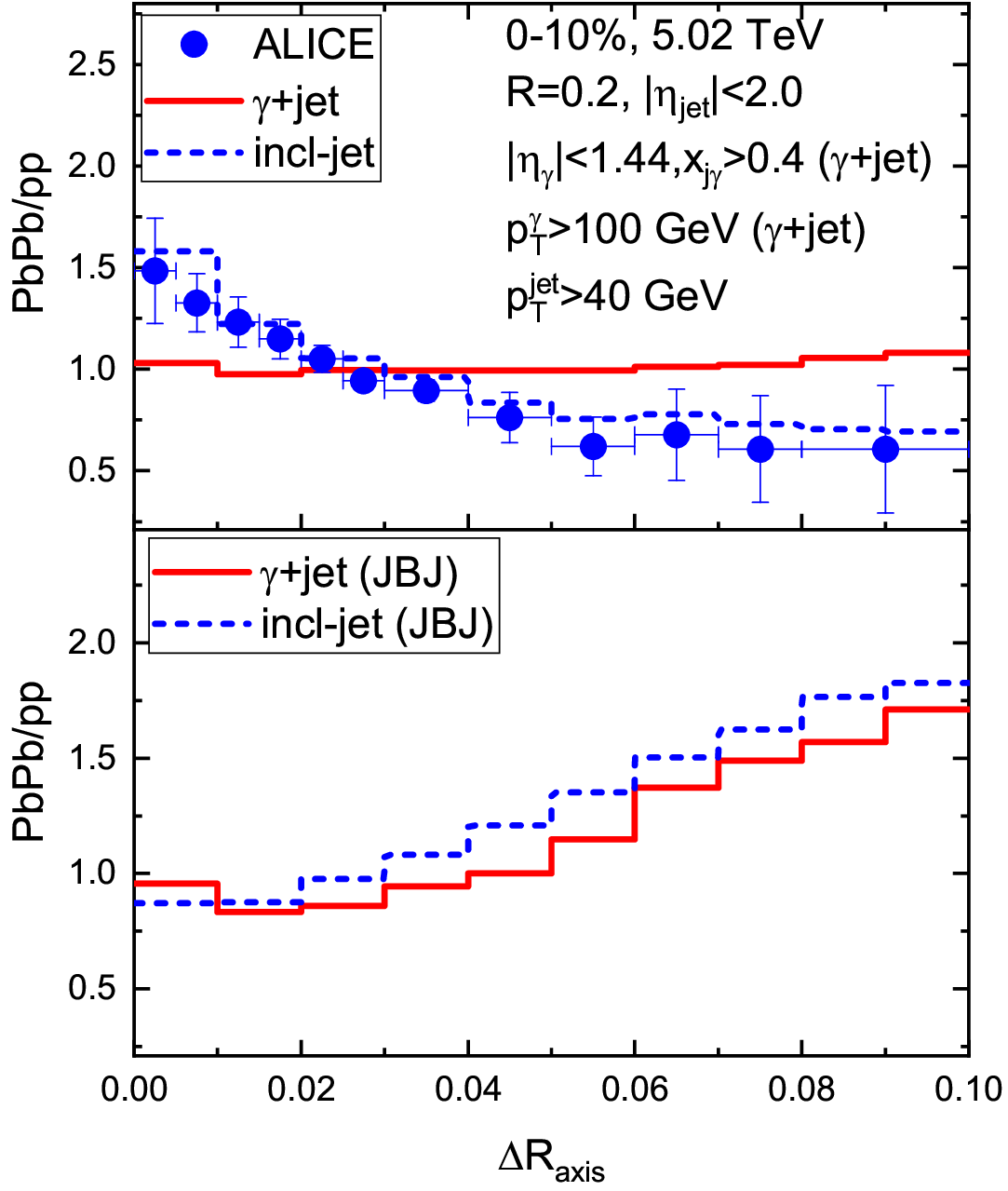}
\vspace*{0in}
\caption{(Color online) Medium modification of the angle between jet axes of $\gamma$+jet and inclusive jet in $0-10\%$ PbPb collisions relative to pp at $\sqrt{s_{NN}}=5.02$ TeV. The results are compared with the measurements of the inclusive jet by the ALICE~\cite{ALICE:2023dwg}. We also plot the results using the Jet-by-Jet matching (JBJ) as a comparison in the lower panel.}
\label{fig:gijetRa}
\end{center}
\end{figure}

In Fig.~\ref{fig:gijetRa}, we compare the medium modification of the angle between jet axes, denoted as $\Delta R_{\rm axis}$, for $\gamma$+jets and inclusive jets in $0-10\%$ PbPb collisions at $\sqrt{s_{NN}} = 5.02$ TeV. The PbPb/pp represents the ratio of normalized $\Delta R_{\rm axis}$ distributions in PbPb to pp collisions. The $\Delta R_{\rm axis}$ quantifies the angular distance between the directions of jet axes reconstructed using both the standard and Winner-Take-All (WTA) recombination schemes~\cite{Cal:2019gxa}:

\begin{eqnarray}
\Delta R_{\rm axis} \equiv \sqrt{(y_{\rm axis}^{\rm WTA} - y_{\rm axis}^{\rm Std})^2 + (\phi_{\rm axis}^{\rm WTA} - \phi_{\rm axis}^{\rm Std})^2}
\end{eqnarray}
Here, $y_{\rm axis}$ and $\phi_{\rm axis}$ represent the rapidity and azimuthal angle of the jet axes, respectively. The standard jet axis is derived by summing the momenta of all constituent particles of the anti-$k_T$ jet using a clustering algorithm known as the E-scheme. In contrast, the WTA axis is obtained by reclustering the anti-$k_T$ jets with the WTA recombination scheme~\cite{Bertolini:2013iqa}. This WTA axis is generally aligned with the most energetic constituent within the jets and minimizes the impact of soft particles on the direction of the jet axis. Therefore, the distance between the standard and WTA axes can be used to investigate the role that soft particles play in the evolution of jets in a medium. Recent studies have measured the medium modification of the inclusive jet $\Delta R_{\rm axis}$ in heavy-ion collisions, as reported by the ALICE collaboration~\cite{ALICE:2023dwg}. Experimental results indicate that inclusive jets show narrower $\Delta R_{\rm axis}$ distributions within the range $40 < p_T^{\rm jet} < 60$ GeV in PbPb collisions compared to pp collisions. Default calculations of the $\Delta R_{\rm axis}$ modification for inclusive jets are consistent with the ALICE data. We also assess the nuclear modification of the $\gamma$+jet $\Delta R_{\rm axis}$ in the same collision system, applying different kinematic cuts ($x_{j\gamma} > 0.4$, $p_{T}^{\gamma} > 100$ GeV). Our calculations reveal a slight enhancement in the PbPb/pp ratio of the $\gamma$+jet $\Delta R_{\rm axis}$ distribution in PbPb collisions for $\Delta R_{\rm axis} > 0.06$ compared to pp collisions. It suggests that angular jet broadening within QGP can also be measured through the $\Delta R_{\rm axis}$ distribution of $\gamma$+jets in heavy-ion collisions. In the lower panel of Fig.~\ref{fig:gijetRa}, the Jet-by-Jet matching results exhibit evident broadening without the influence of selection bias, both for $\gamma$+jet and inclusive jet, which indicates that selection bias suppresses the broadening effect of $\Delta R_{\rm axis}$ distribution in nucleus-nucleus collisions. Especially, the influence of selection bias reverses broadening to narrowing in the $\Delta R_{\rm axis}$ medium modification pattern for inclusive jets. Therefore, the differing nuclear modification patterns of $R_g$ and $\Delta R_{\rm axis}$ for inclusive jets and $\gamma$+jets could be of significant interest for future experimental verification. The findings presented in this paper will aid in interpreting recent measurements at LHC and will be beneficial for future studies focusing on the intrinsic modification of jet substructure in heavy-ion collisions.

%%%%%%%%%%%%%%%%%%%%%%%%%%%%%%%%%%%%%%%%%%%%%%%%%%%%%%%%%%%%%%%%%%%%%
\vspace*{0.1in}
\section{Summary}
\label{sec:sum}

In summary, we present a theoretical study on the angular structure of $\gamma$+jets in high-energy nuclear collisions at the LHC. We utilize PYTHIA8 to provide the initial production of $\gamma$+jets and employ a transport approach to simulate the in-medium jet energy loss in nucleus-nucleus collisions. We carry out the medium modification of $\gamma$+jet girth in $0-30\%$ PbPb collisions at $\sqrt{s_{NN}}=5.02$ TeV, which shows a good agreement with the recently reported CMS data.

Furthermore, we investigate the influence of selection bias when choosing different $x_{j\gamma}$ cuts. With the help of the Jet-by-Jet matching method, we explore the connection between selection bias and kinematic requirements in event selection. Importantly, we quantitatively demonstrate that $\gamma$+jets will provide significant advantages to reduce selection bias and can effectively collect jets sufficiently quenched in PbPb collisions compared to the inclusive jets. We also discuss the contributions of medium-induced gluon radiation and medium response to the broadening of jet angular substructure in PbPb collisions. Additionally, to make the discussion more general, we also study the modification patterns of $R_g$ and $\Delta R_{\rm axis}$ for inclusive jets and $\gamma$+jets in PbPb collisions, which show slight broadening for $\gamma$+jets but significant narrowing for inclusive jets.

This theoretical study provides new insights into the extensive measurements focusing on intra-jet broadening in heavy-ion collisions~\cite{ALargeIonColliderExperiment:2021mqf, ATLAS:2022vii, ALICE:2018dxf, ALICE:2023dwg, Ehlers:2022dfp, ATLAS:2023hso}, as well as the acoplanarity broadening recently observed for lower $p_T$ jets~\cite{STAR:2023pal, STAR:2023ksv, ALICE:2023qve, ALICE:2023jye}. We anticipate further measurements of $\gamma$+jet substructure in heavy-ion collisions, which may provide critical constraints on current theoretical frameworks regarding jet-medium interactions.

\acknowledgments
This research is supported by the National Natural Science Foundation of China with Project Nos.~12535010. S.W. is supported by the Open Foundation of Key Laboratory of Quark and Lepton Physics (MOE) No. QLPL2023P01 and the Talent Scientific Star-up Foundation of China Three Gorges University (CTGU) No. 2024RCKJ013.

\bibliography{sarefs}

\begin{thebibliography}{118}%
\makeatletter
\providecommand \@ifxundefined [1]{%
 \@ifx{#1\undefined}
}%
\providecommand \@ifnum [1]{%
 \ifnum #1\expandafter \@firstoftwo
 \else \expandafter \@secondoftwo
 \fi
}%
\providecommand \@ifx [1]{%
 \ifx #1\expandafter \@firstoftwo
 \else \expandafter \@secondoftwo
 \fi
}%
\providecommand \natexlab [1]{#1}%
\providecommand \enquote  [1]{``#1''}%
\providecommand \bibnamefont  [1]{#1}%
\providecommand \bibfnamefont [1]{#1}%
\providecommand \citenamefont [1]{#1}%
\providecommand \href@noop [0]{\@secondoftwo}%
\providecommand \href [0]{\begingroup \@sanitize@url \@href}%
\providecommand \@href[1]{\@@startlink{#1}\@@href}%
\providecommand \@@href[1]{\endgroup#1\@@endlink}%
\providecommand \@sanitize@url [0]{\catcode `\\12\catcode `\$12\catcode
  `\&12\catcode `\#12\catcode `\^12\catcode `\_12\catcode `\%12\relax}%
\providecommand \@@startlink[1]{}%
\providecommand \@@endlink[0]{}%
\providecommand \url  [0]{\begingroup\@sanitize@url \@url }%
\providecommand \@url [1]{\endgroup\@href {#1}{\urlprefix }}%
\providecommand \urlprefix  [0]{URL }%
\providecommand \Eprint [0]{\href }%
\providecommand \doibase [0]{http://dx.doi.org/}%
\providecommand \selectlanguage [0]{\@gobble}%
\providecommand \bibinfo  [0]{\@secondoftwo}%
\providecommand \bibfield  [0]{\@secondoftwo}%
\providecommand \translation [1]{[#1]}%
\providecommand \BibitemOpen [0]{}%
\providecommand \bibitemStop [0]{}%
\providecommand \bibitemNoStop [0]{.\EOS\space}%
\providecommand \EOS [0]{\spacefactor3000\relax}%
\providecommand \BibitemShut  [1]{\csname bibitem#1\endcsname}%
\let\auto@bib@innerbib\@empty
%</preamble>
\bibitem [{\citenamefont {Gyulassy}\ \emph {et~al.}(2004)\citenamefont
  {Gyulassy}, \citenamefont {Vitev}, \citenamefont {Wang},\ and\ \citenamefont
  {Zhang}}]{Gyulassy:2003mc}%
  \BibitemOpen
  \bibfield  {author} {\bibinfo {author} {\bibfnamefont {M.}~\bibnamefont
  {Gyulassy}}, \bibinfo {author} {\bibfnamefont {I.}~\bibnamefont {Vitev}},
  \bibinfo {author} {\bibfnamefont {X.-N.}\ \bibnamefont {Wang}}, \ and\
  \bibinfo {author} {\bibfnamefont {B.-W.}\ \bibnamefont {Zhang}},\ }\href
  {\doibase 10.1142/9789812795533_0003} {\ ,\ \bibinfo {pages} {123} (\bibinfo
  {year} {2004})},\ \Eprint {http://arxiv.org/abs/nucl-th/0302077}
  {arXiv:nucl-th/0302077} \BibitemShut {NoStop}%
\bibitem [{\citenamefont {Gyulassy}\ and\ \citenamefont
  {Plumer}(1990)}]{Gyulassy:1990ye}%
  \BibitemOpen
  \bibfield  {author} {\bibinfo {author} {\bibfnamefont {M.}~\bibnamefont
  {Gyulassy}}\ and\ \bibinfo {author} {\bibfnamefont {M.}~\bibnamefont
  {Plumer}},\ }\href {\doibase 10.1016/0370-2693(90)91409-5} {\bibfield
  {journal} {\bibinfo  {journal} {Phys. Lett. B}\ }\textbf {\bibinfo {volume}
  {243}},\ \bibinfo {pages} {432} (\bibinfo {year} {1990})}\BibitemShut
  {NoStop}%
\bibitem [{\citenamefont {Qin}\ and\ \citenamefont {Wang}(2015)}]{Qin:2015srf}%
  \BibitemOpen
  \bibfield  {author} {\bibinfo {author} {\bibfnamefont {G.-Y.}\ \bibnamefont
  {Qin}}\ and\ \bibinfo {author} {\bibfnamefont {X.-N.}\ \bibnamefont {Wang}},\
  }\href {\doibase 10.1142/S0218301315300143} {\bibfield  {journal} {\bibinfo
  {journal} {Int. J. Mod. Phys. E}\ }\textbf {\bibinfo {volume} {24}},\
  \bibinfo {pages} {1530014} (\bibinfo {year} {2015})},\ \Eprint
  {http://arxiv.org/abs/1511.00790} {arXiv:1511.00790 [hep-ph]} \BibitemShut
  {NoStop}%
\bibitem [{\citenamefont {Vitev}\ \emph {et~al.}(2008)\citenamefont {Vitev},
  \citenamefont {Wicks},\ and\ \citenamefont {Zhang}}]{Vitev:2008rz}%
  \BibitemOpen
  \bibfield  {author} {\bibinfo {author} {\bibfnamefont {I.}~\bibnamefont
  {Vitev}}, \bibinfo {author} {\bibfnamefont {S.}~\bibnamefont {Wicks}}, \ and\
  \bibinfo {author} {\bibfnamefont {B.-W.}\ \bibnamefont {Zhang}},\ }\href
  {\doibase 10.1088/1126-6708/2008/11/093} {\bibfield  {journal} {\bibinfo
  {journal} {JHEP}\ }\textbf {\bibinfo {volume} {11}},\ \bibinfo {pages} {093}
  (\bibinfo {year} {2008})},\ \Eprint {http://arxiv.org/abs/0810.2807}
  {arXiv:0810.2807 [hep-ph]} \BibitemShut {NoStop}%
\bibitem [{\citenamefont {Armesto}\ \emph {et~al.}(2008)\citenamefont
  {Armesto}, \citenamefont {Borghini}, \citenamefont {Jeon},\ and\
  \citenamefont {Wiedemann}}]{Proceedings:2007ctk}%
  \BibitemOpen
  \bibinfo {editor} {\bibfnamefont {N.}~\bibnamefont {Armesto}}, \bibinfo
  {editor} {\bibfnamefont {N.}~\bibnamefont {Borghini}}, \bibinfo {editor}
  {\bibfnamefont {S.}~\bibnamefont {Jeon}}, \ and\ \bibinfo {editor}
  {\bibfnamefont {U.~A.}\ \bibnamefont {Wiedemann}},\ eds.,\ \href {\doibase
  10.1088/0954-3899/35/5/054001} {\emph {\bibinfo {title} {{Proceedings,
  Workshop on Heavy Ion Collisions at the LHC: Last Call for Predictions}:
  {Geneva, Switzerland, May 14 - June 8, 2007}}}},\ Vol.~\bibinfo {volume}
  {35}\ (\bibinfo {year} {2008})\ \Eprint {http://arxiv.org/abs/0711.0974}
  {arXiv:0711.0974 [hep-ph]} \BibitemShut {NoStop}%
\bibitem [{\citenamefont {Vitev}\ and\ \citenamefont
  {Zhang}(2010)}]{Vitev:2009rd}%
  \BibitemOpen
  \bibfield  {author} {\bibinfo {author} {\bibfnamefont {I.}~\bibnamefont
  {Vitev}}\ and\ \bibinfo {author} {\bibfnamefont {B.-W.}\ \bibnamefont
  {Zhang}},\ }\href {\doibase 10.1103/PhysRevLett.104.132001} {\bibfield
  {journal} {\bibinfo  {journal} {Phys. Rev. Lett.}\ }\textbf {\bibinfo
  {volume} {104}},\ \bibinfo {pages} {132001} (\bibinfo {year} {2010})},\
  \Eprint {http://arxiv.org/abs/0910.1090} {arXiv:0910.1090 [hep-ph]}
  \BibitemShut {NoStop}%
\bibitem [{\citenamefont {Casalderrey-Solana}\ \emph
  {et~al.}(2014)\citenamefont {Casalderrey-Solana}, \citenamefont {Gulhan},
  \citenamefont {Milhano}, \citenamefont {Pablos},\ and\ \citenamefont
  {Rajagopal}}]{Casalderrey-Solana:2014bpa}%
  \BibitemOpen
  \bibfield  {author} {\bibinfo {author} {\bibfnamefont {J.}~\bibnamefont
  {Casalderrey-Solana}}, \bibinfo {author} {\bibfnamefont {D.~C.}\ \bibnamefont
  {Gulhan}}, \bibinfo {author} {\bibfnamefont {J.~G.}\ \bibnamefont {Milhano}},
  \bibinfo {author} {\bibfnamefont {D.}~\bibnamefont {Pablos}}, \ and\ \bibinfo
  {author} {\bibfnamefont {K.}~\bibnamefont {Rajagopal}},\ }\href {\doibase
  10.1007/JHEP09(2015)175} {\bibfield  {journal} {\bibinfo  {journal} {JHEP}\
  }\textbf {\bibinfo {volume} {10}},\ \bibinfo {pages} {019} (\bibinfo {year}
  {2014})},\ \bibinfo {note} {[Erratum: JHEP 09, 175 (2015)]},\ \Eprint
  {http://arxiv.org/abs/1405.3864} {arXiv:1405.3864 [hep-ph]} \BibitemShut
  {NoStop}%
\bibitem [{\citenamefont {Gyulassy}\ and\ \citenamefont
  {Wang}(1994)}]{Gyulassy:1993hr}%
  \BibitemOpen
  \bibfield  {author} {\bibinfo {author} {\bibfnamefont {M.}~\bibnamefont
  {Gyulassy}}\ and\ \bibinfo {author} {\bibfnamefont {X.-n.}\ \bibnamefont
  {Wang}},\ }\href {\doibase 10.1016/0550-3213(94)90079-5} {\bibfield
  {journal} {\bibinfo  {journal} {Nucl. Phys. B}\ }\textbf {\bibinfo {volume}
  {420}},\ \bibinfo {pages} {583} (\bibinfo {year} {1994})},\ \Eprint
  {http://arxiv.org/abs/nucl-th/9306003} {arXiv:nucl-th/9306003} \BibitemShut
  {NoStop}%
\bibitem [{\citenamefont {Wang}\ and\ \citenamefont
  {Guo}(2001)}]{Wang:2001ifa}%
  \BibitemOpen
  \bibfield  {author} {\bibinfo {author} {\bibfnamefont {X.-N.}\ \bibnamefont
  {Wang}}\ and\ \bibinfo {author} {\bibfnamefont {X.-f.}\ \bibnamefont {Guo}},\
  }\href {\doibase 10.1016/S0375-9474(01)01130-7} {\bibfield  {journal}
  {\bibinfo  {journal} {Nucl. Phys. A}\ }\textbf {\bibinfo {volume} {696}},\
  \bibinfo {pages} {788} (\bibinfo {year} {2001})},\ \Eprint
  {http://arxiv.org/abs/hep-ph/0102230} {arXiv:hep-ph/0102230} \BibitemShut
  {NoStop}%
\bibitem [{\citenamefont {Vitev}\ and\ \citenamefont
  {Zhang}(2008)}]{Vitev:2008vk}%
  \BibitemOpen
  \bibfield  {author} {\bibinfo {author} {\bibfnamefont {I.}~\bibnamefont
  {Vitev}}\ and\ \bibinfo {author} {\bibfnamefont {B.-W.}\ \bibnamefont
  {Zhang}},\ }\href {\doibase 10.1016/j.physletb.2008.10.019} {\bibfield
  {journal} {\bibinfo  {journal} {Phys. Lett. B}\ }\textbf {\bibinfo {volume}
  {669}},\ \bibinfo {pages} {337} (\bibinfo {year} {2008})},\ \Eprint
  {http://arxiv.org/abs/0804.3805} {arXiv:0804.3805 [hep-ph]} \BibitemShut
  {NoStop}%
\bibitem [{\citenamefont {Connors}\ \emph {et~al.}(2018)\citenamefont
  {Connors}, \citenamefont {Nattrass}, \citenamefont {Reed},\ and\
  \citenamefont {Salur}}]{Connors:2017ptx}%
  \BibitemOpen
  \bibfield  {author} {\bibinfo {author} {\bibfnamefont {M.}~\bibnamefont
  {Connors}}, \bibinfo {author} {\bibfnamefont {C.}~\bibnamefont {Nattrass}},
  \bibinfo {author} {\bibfnamefont {R.}~\bibnamefont {Reed}}, \ and\ \bibinfo
  {author} {\bibfnamefont {S.}~\bibnamefont {Salur}},\ }\href {\doibase
  10.1103/RevModPhys.90.025005} {\bibfield  {journal} {\bibinfo  {journal}
  {Rev. Mod. Phys.}\ }\textbf {\bibinfo {volume} {90}},\ \bibinfo {pages}
  {025005} (\bibinfo {year} {2018})},\ \Eprint
  {http://arxiv.org/abs/1705.01974} {arXiv:1705.01974 [nucl-ex]} \BibitemShut
  {NoStop}%
\bibitem [{\citenamefont {Andrews}\ \emph {et~al.}(2020)\citenamefont {Andrews}
  \emph {et~al.}}]{Andrews:2018jcm}%
  \BibitemOpen
  \bibfield  {author} {\bibinfo {author} {\bibfnamefont {H.~A.}\ \bibnamefont
  {Andrews}} \emph {et~al.},\ }\href {\doibase 10.1088/1361-6471/ab7cbc}
  {\bibfield  {journal} {\bibinfo  {journal} {J. Phys. G}\ }\textbf {\bibinfo
  {volume} {47}},\ \bibinfo {pages} {065102} (\bibinfo {year} {2020})},\
  \Eprint {http://arxiv.org/abs/1808.03689} {arXiv:1808.03689 [hep-ph]}
  \BibitemShut {NoStop}%
\bibitem [{\citenamefont {Cao}\ and\ \citenamefont {Wang}(2021)}]{Cao:2020wlm}%
  \BibitemOpen
  \bibfield  {author} {\bibinfo {author} {\bibfnamefont {S.}~\bibnamefont
  {Cao}}\ and\ \bibinfo {author} {\bibfnamefont {X.-N.}\ \bibnamefont {Wang}},\
  }\href {\doibase 10.1088/1361-6633/abc22b} {\bibfield  {journal} {\bibinfo
  {journal} {Rept. Prog. Phys.}\ }\textbf {\bibinfo {volume} {84}},\ \bibinfo
  {pages} {024301} (\bibinfo {year} {2021})},\ \Eprint
  {http://arxiv.org/abs/2002.04028} {arXiv:2002.04028 [hep-ph]} \BibitemShut
  {NoStop}%
\bibitem [{\citenamefont {Cunqueiro}\ and\ \citenamefont
  {Sickles}(2022)}]{Cunqueiro:2021wls}%
  \BibitemOpen
  \bibfield  {author} {\bibinfo {author} {\bibfnamefont {L.}~\bibnamefont
  {Cunqueiro}}\ and\ \bibinfo {author} {\bibfnamefont {A.~M.}\ \bibnamefont
  {Sickles}},\ }\href {\doibase 10.1016/j.ppnp.2022.103940} {\bibfield
  {journal} {\bibinfo  {journal} {Prog. Part. Nucl. Phys.}\ }\textbf {\bibinfo
  {volume} {124}},\ \bibinfo {pages} {103940} (\bibinfo {year} {2022})},\
  \Eprint {http://arxiv.org/abs/2110.14490} {arXiv:2110.14490 [nucl-ex]}
  \BibitemShut {NoStop}%
\bibitem [{\citenamefont {Apolin{\'a}rio}\ \emph {et~al.}(2022)\citenamefont
  {Apolin{\'a}rio}, \citenamefont {Lee},\ and\ \citenamefont
  {Winn}}]{Apolinario:2022vzg}%
  \BibitemOpen
  \bibfield  {author} {\bibinfo {author} {\bibfnamefont {L.}~\bibnamefont
  {Apolin{\'a}rio}}, \bibinfo {author} {\bibfnamefont {Y.-J.}\ \bibnamefont
  {Lee}}, \ and\ \bibinfo {author} {\bibfnamefont {M.}~\bibnamefont {Winn}},\
  }\href {\doibase 10.1016/j.ppnp.2022.103990} {\bibfield  {journal} {\bibinfo
  {journal} {Prog. Part. Nucl. Phys.}\ }\textbf {\bibinfo {volume} {127}},\
  \bibinfo {pages} {103990} (\bibinfo {year} {2022})},\ \Eprint
  {http://arxiv.org/abs/2203.16352} {arXiv:2203.16352 [hep-ph]} \BibitemShut
  {NoStop}%
\bibitem [{\citenamefont {Sorensen}\ \emph {et~al.}(2024)\citenamefont
  {Sorensen} \emph {et~al.}}]{Sorensen:2023zkk}%
  \BibitemOpen
  \bibfield  {author} {\bibinfo {author} {\bibfnamefont {A.}~\bibnamefont
  {Sorensen}} \emph {et~al.},\ }\href {\doibase 10.1016/j.ppnp.2023.104080}
  {\bibfield  {journal} {\bibinfo  {journal} {Prog. Part. Nucl. Phys.}\
  }\textbf {\bibinfo {volume} {134}},\ \bibinfo {pages} {104080} (\bibinfo
  {year} {2024})},\ \Eprint {http://arxiv.org/abs/2301.13253} {arXiv:2301.13253
  [nucl-th]} \BibitemShut {NoStop}%
\bibitem [{\citenamefont {Zhao}\ \emph {et~al.}(2020)\citenamefont {Zhao},
  \citenamefont {Ko}, \citenamefont {Liu}, \citenamefont {Qin},\ and\
  \citenamefont {Song}}]{Zhao:2020wcd}%
  \BibitemOpen
  \bibfield  {author} {\bibinfo {author} {\bibfnamefont {W.}~\bibnamefont
  {Zhao}}, \bibinfo {author} {\bibfnamefont {C.~M.}\ \bibnamefont {Ko}},
  \bibinfo {author} {\bibfnamefont {Y.-X.}\ \bibnamefont {Liu}}, \bibinfo
  {author} {\bibfnamefont {G.-Y.}\ \bibnamefont {Qin}}, \ and\ \bibinfo
  {author} {\bibfnamefont {H.}~\bibnamefont {Song}},\ }\href {\doibase
  10.1103/PhysRevLett.125.072301} {\bibfield  {journal} {\bibinfo  {journal}
  {Phys. Rev. Lett.}\ }\textbf {\bibinfo {volume} {125}},\ \bibinfo {pages}
  {072301} (\bibinfo {year} {2020})},\ \Eprint
  {http://arxiv.org/abs/1911.00826} {arXiv:1911.00826 [nucl-th]} \BibitemShut
  {NoStop}%
\bibitem [{\citenamefont {Xie}\ \emph {et~al.}(2024)\citenamefont {Xie},
  \citenamefont {Han}, \citenamefont {Wang}, \citenamefont {Zhang},\ and\
  \citenamefont {Zhang}}]{Xie:2024xbn}%
  \BibitemOpen
  \bibfield  {author} {\bibinfo {author} {\bibfnamefont {M.}~\bibnamefont
  {Xie}}, \bibinfo {author} {\bibfnamefont {Q.-F.}\ \bibnamefont {Han}},
  \bibinfo {author} {\bibfnamefont {E.-K.}\ \bibnamefont {Wang}}, \bibinfo
  {author} {\bibfnamefont {B.-W.}\ \bibnamefont {Zhang}}, \ and\ \bibinfo
  {author} {\bibfnamefont {H.-Z.}\ \bibnamefont {Zhang}},\ }\href {\doibase
  10.1007/s41365-024-01492-4} {\bibfield  {journal} {\bibinfo  {journal} {Nucl.
  Sci. Tech.}\ }\textbf {\bibinfo {volume} {35}},\ \bibinfo {pages} {125}
  (\bibinfo {year} {2024})},\ \Eprint {http://arxiv.org/abs/2409.18773}
  {arXiv:2409.18773 [hep-ph]} \BibitemShut {NoStop}%
\bibitem [{\citenamefont {Zhang}\ \emph
  {et~al.}(2022{\natexlab{a}})\citenamefont {Zhang}, \citenamefont {Xiao},
  \citenamefont {Kang},\ and\ \citenamefont {Zhang}}]{Zhang:2021xib}%
  \BibitemOpen
  \bibfield  {author} {\bibinfo {author} {\bibfnamefont {H.-X.}\ \bibnamefont
  {Zhang}}, \bibinfo {author} {\bibfnamefont {Y.-X.}\ \bibnamefont {Xiao}},
  \bibinfo {author} {\bibfnamefont {J.-W.}\ \bibnamefont {Kang}}, \ and\
  \bibinfo {author} {\bibfnamefont {B.-W.}\ \bibnamefont {Zhang}},\ }\href
  {\doibase 10.1007/s41365-022-01129-4} {\bibfield  {journal} {\bibinfo
  {journal} {Nucl. Sci. Tech.}\ }\textbf {\bibinfo {volume} {33}},\ \bibinfo
  {pages} {150} (\bibinfo {year} {2022}{\natexlab{a}})},\ \Eprint
  {http://arxiv.org/abs/2102.11792} {arXiv:2102.11792 [hep-ph]} \BibitemShut
  {NoStop}%
\bibitem [{\citenamefont {Tang}\ \emph {et~al.}(2020)\citenamefont {Tang},
  \citenamefont {Tang}, \citenamefont {Zha}, \citenamefont {Zha}, \citenamefont
  {Zhang},\ and\ \citenamefont {Zhang}}]{Tang:2020ame}%
  \BibitemOpen
  \bibfield  {author} {\bibinfo {author} {\bibfnamefont {Z.}~\bibnamefont
  {Tang}}, \bibinfo {author} {\bibfnamefont {Z.-B.}\ \bibnamefont {Tang}},
  \bibinfo {author} {\bibfnamefont {W.}~\bibnamefont {Zha}}, \bibinfo {author}
  {\bibfnamefont {W.-M.}\ \bibnamefont {Zha}}, \bibinfo {author} {\bibfnamefont
  {Y.}~\bibnamefont {Zhang}}, \ and\ \bibinfo {author} {\bibfnamefont {Y.-F.}\
  \bibnamefont {Zhang}},\ }\href {\doibase 10.1007/s41365-020-00785-8}
  {\bibfield  {journal} {\bibinfo  {journal} {Nucl. Sci. Tech.}\ }\textbf
  {\bibinfo {volume} {31}},\ \bibinfo {pages} {81} (\bibinfo {year} {2020})},\
  \Eprint {http://arxiv.org/abs/2105.11656} {arXiv:2105.11656 [nucl-ex]}
  \BibitemShut {NoStop}%
\bibitem [{\citenamefont {Shen}\ and\ \citenamefont
  {Yan}(2020)}]{Shen:2020mgh}%
  \BibitemOpen
  \bibfield  {author} {\bibinfo {author} {\bibfnamefont {C.}~\bibnamefont
  {Shen}}\ and\ \bibinfo {author} {\bibfnamefont {L.}~\bibnamefont {Yan}},\
  }\href {\doibase 10.1007/s41365-020-00829-z} {\bibfield  {journal} {\bibinfo
  {journal} {Nucl. Sci. Tech.}\ }\textbf {\bibinfo {volume} {31}},\ \bibinfo
  {pages} {122} (\bibinfo {year} {2020})},\ \Eprint
  {http://arxiv.org/abs/2010.12377} {arXiv:2010.12377 [nucl-th]} \BibitemShut
  {NoStop}%
\bibitem [{\citenamefont {Chen}\ \emph {et~al.}(2025)\citenamefont {Chen},
  \citenamefont {Chen}, \citenamefont {Guo}, \citenamefont {Ma}, \citenamefont
  {Shen}, \citenamefont {Shou}, \citenamefont {Shou}, \citenamefont {Wang},
  \citenamefont {Wu},\ and\ \citenamefont {Zou}}]{Chen:2024eaq}%
  \BibitemOpen
  \bibfield  {author} {\bibinfo {author} {\bibfnamefont {J.-H.}\ \bibnamefont
  {Chen}}, \bibinfo {author} {\bibfnamefont {J.}~\bibnamefont {Chen}}, \bibinfo
  {author} {\bibfnamefont {F.-K.}\ \bibnamefont {Guo}}, \bibinfo {author}
  {\bibfnamefont {Y.-G.}\ \bibnamefont {Ma}}, \bibinfo {author} {\bibfnamefont
  {C.-P.}\ \bibnamefont {Shen}}, \bibinfo {author} {\bibfnamefont {Q.-Y.}\
  \bibnamefont {Shou}}, \bibinfo {author} {\bibfnamefont {Q.}~\bibnamefont
  {Shou}}, \bibinfo {author} {\bibfnamefont {Q.}~\bibnamefont {Wang}}, \bibinfo
  {author} {\bibfnamefont {J.-J.}\ \bibnamefont {Wu}}, \ and\ \bibinfo {author}
  {\bibfnamefont {B.-S.}\ \bibnamefont {Zou}},\ }\href {\doibase
  10.1007/s41365-025-01664-w} {\bibfield  {journal} {\bibinfo  {journal} {Nucl.
  Sci. Tech.}\ }\textbf {\bibinfo {volume} {36}},\ \bibinfo {pages} {55}
  (\bibinfo {year} {2025})},\ \Eprint {http://arxiv.org/abs/2411.18257}
  {arXiv:2411.18257 [hep-ph]} \BibitemShut {NoStop}%
\bibitem [{\citenamefont {Chen}\ \emph {et~al.}(2024)\citenamefont {Chen} \emph
  {et~al.}}]{Chen:2024aom}%
  \BibitemOpen
  \bibfield  {author} {\bibinfo {author} {\bibfnamefont {J.}~\bibnamefont
  {Chen}} \emph {et~al.},\ }\href {\doibase 10.1007/s41365-024-01591-2}
  {\bibfield  {journal} {\bibinfo  {journal} {Nucl. Sci. Tech.}\ }\textbf
  {\bibinfo {volume} {35}},\ \bibinfo {pages} {214} (\bibinfo {year} {2024})},\
  \Eprint {http://arxiv.org/abs/2407.02935} {arXiv:2407.02935 [nucl-ex]}
  \BibitemShut {NoStop}%
\bibitem [{\citenamefont {Zhao}\ \emph {et~al.}(2024)\citenamefont {Zhao},
  \citenamefont {Chen}, \citenamefont {Huang},\ and\ \citenamefont
  {Ma}}]{Zhao:2022dac}%
  \BibitemOpen
  \bibfield  {author} {\bibinfo {author} {\bibfnamefont {J.}~\bibnamefont
  {Zhao}}, \bibinfo {author} {\bibfnamefont {J.-H.}\ \bibnamefont {Chen}},
  \bibinfo {author} {\bibfnamefont {X.-G.}\ \bibnamefont {Huang}}, \ and\
  \bibinfo {author} {\bibfnamefont {Y.-G.}\ \bibnamefont {Ma}},\ }\href
  {\doibase 10.1007/s41365-024-01374-9} {\bibfield  {journal} {\bibinfo
  {journal} {Nucl. Sci. Tech.}\ }\textbf {\bibinfo {volume} {35}},\ \bibinfo
  {pages} {20} (\bibinfo {year} {2024})},\ \Eprint
  {http://arxiv.org/abs/2211.03968} {arXiv:2211.03968 [nucl-th]} \BibitemShut
  {NoStop}%
\bibitem [{\citenamefont {Shou}\ \emph {et~al.}(2024)\citenamefont {Shou} \emph
  {et~al.}}]{Shou:2024uga}%
  \BibitemOpen
  \bibfield  {author} {\bibinfo {author} {\bibfnamefont {Q.-Y.}\ \bibnamefont
  {Shou}} \emph {et~al.},\ }\href {\doibase 10.1007/s41365-024-01583-2}
  {\bibfield  {journal} {\bibinfo  {journal} {Nucl. Sci. Tech.}\ }\textbf
  {\bibinfo {volume} {35}},\ \bibinfo {pages} {219} (\bibinfo {year} {2024})},\
  \Eprint {http://arxiv.org/abs/2409.17964} {arXiv:2409.17964 [nucl-ex]}
  \BibitemShut {NoStop}%
\bibitem [{\citenamefont {Baier}\ \emph
  {et~al.}(1997{\natexlab{a}})\citenamefont {Baier}, \citenamefont
  {Dokshitzer}, \citenamefont {Mueller}, \citenamefont {Peigne},\ and\
  \citenamefont {Schiff}}]{Baier:1996kr}%
  \BibitemOpen
  \bibfield  {author} {\bibinfo {author} {\bibfnamefont {R.}~\bibnamefont
  {Baier}}, \bibinfo {author} {\bibfnamefont {Y.~L.}\ \bibnamefont
  {Dokshitzer}}, \bibinfo {author} {\bibfnamefont {A.~H.}\ \bibnamefont
  {Mueller}}, \bibinfo {author} {\bibfnamefont {S.}~\bibnamefont {Peigne}}, \
  and\ \bibinfo {author} {\bibfnamefont {D.}~\bibnamefont {Schiff}},\ }\href
  {\doibase 10.1016/S0550-3213(96)00553-6} {\bibfield  {journal} {\bibinfo
  {journal} {Nucl. Phys. B}\ }\textbf {\bibinfo {volume} {483}},\ \bibinfo
  {pages} {291} (\bibinfo {year} {1997}{\natexlab{a}})},\ \Eprint
  {http://arxiv.org/abs/hep-ph/9607355} {arXiv:hep-ph/9607355} \BibitemShut
  {NoStop}%
\bibitem [{\citenamefont {Baier}\ \emph
  {et~al.}(1997{\natexlab{b}})\citenamefont {Baier}, \citenamefont
  {Dokshitzer}, \citenamefont {Mueller}, \citenamefont {Peigne},\ and\
  \citenamefont {Schiff}}]{Baier:1996sk}%
  \BibitemOpen
  \bibfield  {author} {\bibinfo {author} {\bibfnamefont {R.}~\bibnamefont
  {Baier}}, \bibinfo {author} {\bibfnamefont {Y.~L.}\ \bibnamefont
  {Dokshitzer}}, \bibinfo {author} {\bibfnamefont {A.~H.}\ \bibnamefont
  {Mueller}}, \bibinfo {author} {\bibfnamefont {S.}~\bibnamefont {Peigne}}, \
  and\ \bibinfo {author} {\bibfnamefont {D.}~\bibnamefont {Schiff}},\ }\href
  {\doibase 10.1016/S0550-3213(96)00581-0} {\bibfield  {journal} {\bibinfo
  {journal} {Nucl. Phys. B}\ }\textbf {\bibinfo {volume} {484}},\ \bibinfo
  {pages} {265} (\bibinfo {year} {1997}{\natexlab{b}})},\ \Eprint
  {http://arxiv.org/abs/hep-ph/9608322} {arXiv:hep-ph/9608322} \BibitemShut
  {NoStop}%
\bibitem [{\citenamefont {Kunnawalkam~Elayavalli}\ and\ \citenamefont
  {Zapp}(2017)}]{KunnawalkamElayavalli:2017hxo}%
  \BibitemOpen
  \bibfield  {author} {\bibinfo {author} {\bibfnamefont {R.}~\bibnamefont
  {Kunnawalkam~Elayavalli}}\ and\ \bibinfo {author} {\bibfnamefont {K.~C.}\
  \bibnamefont {Zapp}},\ }\href {\doibase 10.1007/JHEP07(2017)141} {\bibfield
  {journal} {\bibinfo  {journal} {JHEP}\ }\textbf {\bibinfo {volume} {07}},\
  \bibinfo {pages} {141} (\bibinfo {year} {2017})},\ \Eprint
  {http://arxiv.org/abs/1707.01539} {arXiv:1707.01539 [hep-ph]} \BibitemShut
  {NoStop}%
\bibitem [{\citenamefont {Pablos}(2020)}]{Pablos:2019ngg}%
  \BibitemOpen
  \bibfield  {author} {\bibinfo {author} {\bibfnamefont {D.}~\bibnamefont
  {Pablos}},\ }\href {\doibase 10.1103/PhysRevLett.124.052301} {\bibfield
  {journal} {\bibinfo  {journal} {Phys. Rev. Lett.}\ }\textbf {\bibinfo
  {volume} {124}},\ \bibinfo {pages} {052301} (\bibinfo {year} {2020})},\
  \Eprint {http://arxiv.org/abs/1907.12301} {arXiv:1907.12301 [hep-ph]}
  \BibitemShut {NoStop}%
\bibitem [{\citenamefont {Chen}\ \emph {et~al.}(2020)\citenamefont {Chen},
  \citenamefont {Cao}, \citenamefont {Luo}, \citenamefont {Pang},\ and\
  \citenamefont {Wang}}]{Chen:2020tbl}%
  \BibitemOpen
  \bibfield  {author} {\bibinfo {author} {\bibfnamefont {W.}~\bibnamefont
  {Chen}}, \bibinfo {author} {\bibfnamefont {S.}~\bibnamefont {Cao}}, \bibinfo
  {author} {\bibfnamefont {T.}~\bibnamefont {Luo}}, \bibinfo {author}
  {\bibfnamefont {L.-G.}\ \bibnamefont {Pang}}, \ and\ \bibinfo {author}
  {\bibfnamefont {X.-N.}\ \bibnamefont {Wang}},\ }\href {\doibase
  10.1016/j.physletb.2020.135783} {\bibfield  {journal} {\bibinfo  {journal}
  {Phys. Lett. B}\ }\textbf {\bibinfo {volume} {810}},\ \bibinfo {pages}
  {135783} (\bibinfo {year} {2020})},\ \Eprint
  {http://arxiv.org/abs/2005.09678} {arXiv:2005.09678 [hep-ph]} \BibitemShut
  {NoStop}%
\bibitem [{\citenamefont {Casalderrey-Solana}\ \emph
  {et~al.}(2021)\citenamefont {Casalderrey-Solana}, \citenamefont {Milhano},
  \citenamefont {Pablos}, \citenamefont {Rajagopal},\ and\ \citenamefont
  {Yao}}]{Casalderrey-Solana:2020rsj}%
  \BibitemOpen
  \bibfield  {author} {\bibinfo {author} {\bibfnamefont {J.}~\bibnamefont
  {Casalderrey-Solana}}, \bibinfo {author} {\bibfnamefont {J.~G.}\ \bibnamefont
  {Milhano}}, \bibinfo {author} {\bibfnamefont {D.}~\bibnamefont {Pablos}},
  \bibinfo {author} {\bibfnamefont {K.}~\bibnamefont {Rajagopal}}, \ and\
  \bibinfo {author} {\bibfnamefont {X.}~\bibnamefont {Yao}},\ }\href {\doibase
  10.1007/JHEP05(2021)230} {\bibfield  {journal} {\bibinfo  {journal} {JHEP}\
  }\textbf {\bibinfo {volume} {05}},\ \bibinfo {pages} {230} (\bibinfo {year}
  {2021})},\ \Eprint {http://arxiv.org/abs/2010.01140} {arXiv:2010.01140
  [hep-ph]} \BibitemShut {NoStop}%
\bibitem [{\citenamefont {He}\ \emph {et~al.}(2019)\citenamefont {He},
  \citenamefont {Cao}, \citenamefont {Chen}, \citenamefont {Luo}, \citenamefont
  {Pang},\ and\ \citenamefont {Wang}}]{He:2018xjv}%
  \BibitemOpen
  \bibfield  {author} {\bibinfo {author} {\bibfnamefont {Y.}~\bibnamefont
  {He}}, \bibinfo {author} {\bibfnamefont {S.}~\bibnamefont {Cao}}, \bibinfo
  {author} {\bibfnamefont {W.}~\bibnamefont {Chen}}, \bibinfo {author}
  {\bibfnamefont {T.}~\bibnamefont {Luo}}, \bibinfo {author} {\bibfnamefont
  {L.-G.}\ \bibnamefont {Pang}}, \ and\ \bibinfo {author} {\bibfnamefont
  {X.-N.}\ \bibnamefont {Wang}},\ }\href {\doibase 10.1103/PhysRevC.99.054911}
  {\bibfield  {journal} {\bibinfo  {journal} {Phys. Rev. C}\ }\textbf {\bibinfo
  {volume} {99}},\ \bibinfo {pages} {054911} (\bibinfo {year} {2019})},\
  \Eprint {http://arxiv.org/abs/1809.02525} {arXiv:1809.02525 [nucl-th]}
  \BibitemShut {NoStop}%
\bibitem [{\citenamefont {Ke}\ and\ \citenamefont {Wang}(2021)}]{Ke:2020clc}%
  \BibitemOpen
  \bibfield  {author} {\bibinfo {author} {\bibfnamefont {W.}~\bibnamefont
  {Ke}}\ and\ \bibinfo {author} {\bibfnamefont {X.-N.}\ \bibnamefont {Wang}},\
  }\href {\doibase 10.1007/JHEP05(2021)041} {\bibfield  {journal} {\bibinfo
  {journal} {JHEP}\ }\textbf {\bibinfo {volume} {05}},\ \bibinfo {pages} {041}
  (\bibinfo {year} {2021})},\ \Eprint {http://arxiv.org/abs/2010.13680}
  {arXiv:2010.13680 [hep-ph]} \BibitemShut {NoStop}%
\bibitem [{\citenamefont {Hulcher}\ \emph {et~al.}(2018)\citenamefont
  {Hulcher}, \citenamefont {Pablos},\ and\ \citenamefont
  {Rajagopal}}]{Hulcher:2017cpt}%
  \BibitemOpen
  \bibfield  {author} {\bibinfo {author} {\bibfnamefont {Z.}~\bibnamefont
  {Hulcher}}, \bibinfo {author} {\bibfnamefont {D.}~\bibnamefont {Pablos}}, \
  and\ \bibinfo {author} {\bibfnamefont {K.}~\bibnamefont {Rajagopal}},\ }\href
  {\doibase 10.1007/JHEP03(2018)010} {\bibfield  {journal} {\bibinfo  {journal}
  {JHEP}\ }\textbf {\bibinfo {volume} {03}},\ \bibinfo {pages} {010} (\bibinfo
  {year} {2018})},\ \Eprint {http://arxiv.org/abs/1707.05245} {arXiv:1707.05245
  [hep-ph]} \BibitemShut {NoStop}%
\bibitem [{\citenamefont {Mehtar-Tani}\ and\ \citenamefont
  {Tywoniuk}(2017)}]{Mehtar-Tani:2016aco}%
  \BibitemOpen
  \bibfield  {author} {\bibinfo {author} {\bibfnamefont {Y.}~\bibnamefont
  {Mehtar-Tani}}\ and\ \bibinfo {author} {\bibfnamefont {K.}~\bibnamefont
  {Tywoniuk}},\ }\href {\doibase 10.1007/JHEP04(2017)125} {\bibfield  {journal}
  {\bibinfo  {journal} {JHEP}\ }\textbf {\bibinfo {volume} {04}},\ \bibinfo
  {pages} {125} (\bibinfo {year} {2017})},\ \Eprint
  {http://arxiv.org/abs/1610.08930} {arXiv:1610.08930 [hep-ph]} \BibitemShut
  {NoStop}%
\bibitem [{\citenamefont {Caucal}\ \emph {et~al.}(2018)\citenamefont {Caucal},
  \citenamefont {Iancu}, \citenamefont {Mueller},\ and\ \citenamefont
  {Soyez}}]{Caucal:2018dla}%
  \BibitemOpen
  \bibfield  {author} {\bibinfo {author} {\bibfnamefont {P.}~\bibnamefont
  {Caucal}}, \bibinfo {author} {\bibfnamefont {E.}~\bibnamefont {Iancu}},
  \bibinfo {author} {\bibfnamefont {A.~H.}\ \bibnamefont {Mueller}}, \ and\
  \bibinfo {author} {\bibfnamefont {G.}~\bibnamefont {Soyez}},\ }\href
  {\doibase 10.1103/PhysRevLett.120.232001} {\bibfield  {journal} {\bibinfo
  {journal} {Phys. Rev. Lett.}\ }\textbf {\bibinfo {volume} {120}},\ \bibinfo
  {pages} {232001} (\bibinfo {year} {2018})},\ \Eprint
  {http://arxiv.org/abs/1801.09703} {arXiv:1801.09703 [hep-ph]} \BibitemShut
  {NoStop}%
\bibitem [{\citenamefont {D'Eramo}\ \emph {et~al.}(2013)\citenamefont
  {D'Eramo}, \citenamefont {Lekaveckas}, \citenamefont {Liu},\ and\
  \citenamefont {Rajagopal}}]{DEramo:2012uzl}%
  \BibitemOpen
  \bibfield  {author} {\bibinfo {author} {\bibfnamefont {F.}~\bibnamefont
  {D'Eramo}}, \bibinfo {author} {\bibfnamefont {M.}~\bibnamefont {Lekaveckas}},
  \bibinfo {author} {\bibfnamefont {H.}~\bibnamefont {Liu}}, \ and\ \bibinfo
  {author} {\bibfnamefont {K.}~\bibnamefont {Rajagopal}},\ }\href {\doibase
  10.1007/JHEP05(2013)031} {\bibfield  {journal} {\bibinfo  {journal} {JHEP}\
  }\textbf {\bibinfo {volume} {05}},\ \bibinfo {pages} {031} (\bibinfo {year}
  {2013})},\ \Eprint {http://arxiv.org/abs/1211.1922} {arXiv:1211.1922
  [hep-ph]} \BibitemShut {NoStop}%
\bibitem [{\citenamefont {D'Eramo}\ \emph {et~al.}(2019)\citenamefont
  {D'Eramo}, \citenamefont {Rajagopal},\ and\ \citenamefont
  {Yin}}]{DEramo:2018eoy}%
  \BibitemOpen
  \bibfield  {author} {\bibinfo {author} {\bibfnamefont {F.}~\bibnamefont
  {D'Eramo}}, \bibinfo {author} {\bibfnamefont {K.}~\bibnamefont {Rajagopal}},
  \ and\ \bibinfo {author} {\bibfnamefont {Y.}~\bibnamefont {Yin}},\ }\href
  {\doibase 10.1007/JHEP01(2019)172} {\bibfield  {journal} {\bibinfo  {journal}
  {JHEP}\ }\textbf {\bibinfo {volume} {01}},\ \bibinfo {pages} {172} (\bibinfo
  {year} {2019})},\ \Eprint {http://arxiv.org/abs/1808.03250} {arXiv:1808.03250
  [hep-ph]} \BibitemShut {NoStop}%
\bibitem [{\citenamefont {Hayrapetyan}\ \emph
  {et~al.}(2025{\natexlab{a}})\citenamefont {Hayrapetyan} \emph
  {et~al.}}]{CMS:2024krd}%
  \BibitemOpen
  \bibfield  {author} {\bibinfo {author} {\bibfnamefont {A.}~\bibnamefont
  {Hayrapetyan}} \emph {et~al.} (\bibinfo {collaboration} {CMS}),\ }\href
  {\doibase 10.1016/j.physrep.2024.11.007} {\bibfield  {journal} {\bibinfo
  {journal} {Phys. Rept.}\ }\textbf {\bibinfo {volume} {1115}},\ \bibinfo
  {pages} {219} (\bibinfo {year} {2025}{\natexlab{a}})},\ \Eprint
  {http://arxiv.org/abs/2405.10785} {arXiv:2405.10785 [nucl-ex]} \BibitemShut
  {NoStop}%
\bibitem [{\citenamefont {Arslandok}\ \emph {et~al.}(2023)\citenamefont
  {Arslandok} \emph {et~al.}}]{Arslandok:2023utm}%
  \BibitemOpen
  \bibfield  {author} {\bibinfo {author} {\bibfnamefont {M.}~\bibnamefont
  {Arslandok}} \emph {et~al.},\ }\href@noop {} {\  (\bibinfo {year} {2023})},\
  \Eprint {http://arxiv.org/abs/2303.17254} {arXiv:2303.17254 [nucl-ex]}
  \BibitemShut {NoStop}%
\bibitem [{\citenamefont {Marzani}\ \emph {et~al.}(2019)\citenamefont
  {Marzani}, \citenamefont {Soyez},\ and\ \citenamefont
  {Spannowsky}}]{Marzani:2019hun}%
  \BibitemOpen
  \bibfield  {author} {\bibinfo {author} {\bibfnamefont {S.}~\bibnamefont
  {Marzani}}, \bibinfo {author} {\bibfnamefont {G.}~\bibnamefont {Soyez}}, \
  and\ \bibinfo {author} {\bibfnamefont {M.}~\bibnamefont {Spannowsky}},\
  }\href {\doibase 10.1007/978-3-030-15709-8} {\emph {\bibinfo {title}
  {{Looking inside jets: an introduction to jet substructure and boosted-object
  phenomenology}}}},\ Vol.\ \bibinfo {volume} {958}\ (\bibinfo  {publisher}
  {Springer},\ \bibinfo {year} {2019})\ \Eprint
  {http://arxiv.org/abs/1901.10342} {arXiv:1901.10342 [hep-ph]} \BibitemShut
  {NoStop}%
\bibitem [{\citenamefont {Ringer}\ \emph {et~al.}(2020)\citenamefont {Ringer},
  \citenamefont {Xiao},\ and\ \citenamefont {Yuan}}]{Ringer:2019rfk}%
  \BibitemOpen
  \bibfield  {author} {\bibinfo {author} {\bibfnamefont {F.}~\bibnamefont
  {Ringer}}, \bibinfo {author} {\bibfnamefont {B.-W.}\ \bibnamefont {Xiao}}, \
  and\ \bibinfo {author} {\bibfnamefont {F.}~\bibnamefont {Yuan}},\ }\href
  {\doibase 10.1016/j.physletb.2020.135634} {\bibfield  {journal} {\bibinfo
  {journal} {Phys. Lett. B}\ }\textbf {\bibinfo {volume} {808}},\ \bibinfo
  {pages} {135634} (\bibinfo {year} {2020})},\ \Eprint
  {http://arxiv.org/abs/1907.12541} {arXiv:1907.12541 [hep-ph]} \BibitemShut
  {NoStop}%
\bibitem [{\citenamefont {Rajagopal}\ \emph {et~al.}(2016)\citenamefont
  {Rajagopal}, \citenamefont {Sadofyev},\ and\ \citenamefont {van~der
  Schee}}]{Rajagopal:2016uip}%
  \BibitemOpen
  \bibfield  {author} {\bibinfo {author} {\bibfnamefont {K.}~\bibnamefont
  {Rajagopal}}, \bibinfo {author} {\bibfnamefont {A.~V.}\ \bibnamefont
  {Sadofyev}}, \ and\ \bibinfo {author} {\bibfnamefont {W.}~\bibnamefont
  {van~der Schee}},\ }\href {\doibase 10.1103/PhysRevLett.116.211603}
  {\bibfield  {journal} {\bibinfo  {journal} {Phys. Rev. Lett.}\ }\textbf
  {\bibinfo {volume} {116}},\ \bibinfo {pages} {211603} (\bibinfo {year}
  {2016})},\ \Eprint {http://arxiv.org/abs/1602.04187} {arXiv:1602.04187
  [nucl-th]} \BibitemShut {NoStop}%
\bibitem [{\citenamefont {Chien}\ and\ \citenamefont
  {Vitev}(2017)}]{Chien:2016led}%
  \BibitemOpen
  \bibfield  {author} {\bibinfo {author} {\bibfnamefont {Y.-T.}\ \bibnamefont
  {Chien}}\ and\ \bibinfo {author} {\bibfnamefont {I.}~\bibnamefont {Vitev}},\
  }\href {\doibase 10.1103/PhysRevLett.119.112301} {\bibfield  {journal}
  {\bibinfo  {journal} {Phys. Rev. Lett.}\ }\textbf {\bibinfo {volume} {119}},\
  \bibinfo {pages} {112301} (\bibinfo {year} {2017})},\ \Eprint
  {http://arxiv.org/abs/1608.07283} {arXiv:1608.07283 [hep-ph]} \BibitemShut
  {NoStop}%
\bibitem [{\citenamefont {Larkoski}\ \emph {et~al.}(2017)\citenamefont
  {Larkoski}, \citenamefont {Marzani}, \citenamefont {Thaler}, \citenamefont
  {Tripathee},\ and\ \citenamefont {Xue}}]{Larkoski:2017bvj}%
  \BibitemOpen
  \bibfield  {author} {\bibinfo {author} {\bibfnamefont {A.}~\bibnamefont
  {Larkoski}}, \bibinfo {author} {\bibfnamefont {S.}~\bibnamefont {Marzani}},
  \bibinfo {author} {\bibfnamefont {J.}~\bibnamefont {Thaler}}, \bibinfo
  {author} {\bibfnamefont {A.}~\bibnamefont {Tripathee}}, \ and\ \bibinfo
  {author} {\bibfnamefont {W.}~\bibnamefont {Xue}},\ }\href {\doibase
  10.1103/PhysRevLett.119.132003} {\bibfield  {journal} {\bibinfo  {journal}
  {Phys. Rev. Lett.}\ }\textbf {\bibinfo {volume} {119}},\ \bibinfo {pages}
  {132003} (\bibinfo {year} {2017})},\ \Eprint
  {http://arxiv.org/abs/1704.05066} {arXiv:1704.05066 [hep-ph]} \BibitemShut
  {NoStop}%
\bibitem [{\citenamefont {Chang}\ \emph {et~al.}(2018)\citenamefont {Chang},
  \citenamefont {Cao},\ and\ \citenamefont {Qin}}]{Chang:2017gkt}%
  \BibitemOpen
  \bibfield  {author} {\bibinfo {author} {\bibfnamefont {N.-B.}\ \bibnamefont
  {Chang}}, \bibinfo {author} {\bibfnamefont {S.}~\bibnamefont {Cao}}, \ and\
  \bibinfo {author} {\bibfnamefont {G.-Y.}\ \bibnamefont {Qin}},\ }\href
  {\doibase 10.1016/j.physletb.2018.04.019} {\bibfield  {journal} {\bibinfo
  {journal} {Phys. Lett. B}\ }\textbf {\bibinfo {volume} {781}},\ \bibinfo
  {pages} {423} (\bibinfo {year} {2018})},\ \Eprint
  {http://arxiv.org/abs/1707.03767} {arXiv:1707.03767 [hep-ph]} \BibitemShut
  {NoStop}%
\bibitem [{\citenamefont {Casalderrey-Solana}\ \emph
  {et~al.}(2017)\citenamefont {Casalderrey-Solana}, \citenamefont {Gulhan},
  \citenamefont {Milhano}, \citenamefont {Pablos},\ and\ \citenamefont
  {Rajagopal}}]{Casalderrey-Solana:2016jvj}%
  \BibitemOpen
  \bibfield  {author} {\bibinfo {author} {\bibfnamefont {J.}~\bibnamefont
  {Casalderrey-Solana}}, \bibinfo {author} {\bibfnamefont {D.}~\bibnamefont
  {Gulhan}}, \bibinfo {author} {\bibfnamefont {G.}~\bibnamefont {Milhano}},
  \bibinfo {author} {\bibfnamefont {D.}~\bibnamefont {Pablos}}, \ and\ \bibinfo
  {author} {\bibfnamefont {K.}~\bibnamefont {Rajagopal}},\ }\href {\doibase
  10.1007/JHEP03(2017)135} {\bibfield  {journal} {\bibinfo  {journal} {JHEP}\
  }\textbf {\bibinfo {volume} {03}},\ \bibinfo {pages} {135} (\bibinfo {year}
  {2017})},\ \Eprint {http://arxiv.org/abs/1609.05842} {arXiv:1609.05842
  [hep-ph]} \BibitemShut {NoStop}%
\bibitem [{\citenamefont {Kang}\ \emph
  {et~al.}(2025{\natexlab{a}})\citenamefont {Kang}, \citenamefont {Wang},
  \citenamefont {Wang},\ and\ \citenamefont {Zhang}}]{Kang:2023ycg}%
  \BibitemOpen
  \bibfield  {author} {\bibinfo {author} {\bibfnamefont {J.-W.}\ \bibnamefont
  {Kang}}, \bibinfo {author} {\bibfnamefont {S.}~\bibnamefont {Wang}}, \bibinfo
  {author} {\bibfnamefont {L.}~\bibnamefont {Wang}}, \ and\ \bibinfo {author}
  {\bibfnamefont {B.-W.}\ \bibnamefont {Zhang}},\ }\href {\doibase
  10.1103/PhysRevC.111.054905} {\bibfield  {journal} {\bibinfo  {journal}
  {Phys. Rev. C}\ }\textbf {\bibinfo {volume} {111}},\ \bibinfo {pages}
  {054905} (\bibinfo {year} {2025}{\natexlab{a}})},\ \Eprint
  {http://arxiv.org/abs/2312.15518} {arXiv:2312.15518 [hep-ph]} \BibitemShut
  {NoStop}%
\bibitem [{\citenamefont {Kang}\ \emph
  {et~al.}(2025{\natexlab{b}})\citenamefont {Kang}, \citenamefont {Wang},
  \citenamefont {Dai}, \citenamefont {Wang},\ and\ \citenamefont
  {Zhang}}]{Kang:2023qxb}%
  \BibitemOpen
  \bibfield  {author} {\bibinfo {author} {\bibfnamefont {J.-W.}\ \bibnamefont
  {Kang}}, \bibinfo {author} {\bibfnamefont {L.}~\bibnamefont {Wang}}, \bibinfo
  {author} {\bibfnamefont {W.}~\bibnamefont {Dai}}, \bibinfo {author}
  {\bibfnamefont {S.}~\bibnamefont {Wang}}, \ and\ \bibinfo {author}
  {\bibfnamefont {B.-W.}\ \bibnamefont {Zhang}},\ }\href {\doibase
  10.1103/71sd-7qqb} {\bibfield  {journal} {\bibinfo  {journal} {Phys. Rev. C}\
  }\textbf {\bibinfo {volume} {112}},\ \bibinfo {pages} {034903} (\bibinfo
  {year} {2025}{\natexlab{b}})},\ \Eprint {http://arxiv.org/abs/2304.04649}
  {arXiv:2304.04649 [nucl-th]} \BibitemShut {NoStop}%
\bibitem [{\citenamefont {Tachibana}\ \emph {et~al.}(2024)\citenamefont
  {Tachibana} \emph {et~al.}}]{JETSCAPE:2023hqn}%
  \BibitemOpen
  \bibfield  {author} {\bibinfo {author} {\bibfnamefont {Y.}~\bibnamefont
  {Tachibana}} \emph {et~al.} (\bibinfo {collaboration} {JETSCAPE}),\ }\href
  {\doibase 10.1103/PhysRevC.110.044907} {\bibfield  {journal} {\bibinfo
  {journal} {Phys. Rev. C}\ }\textbf {\bibinfo {volume} {110}},\ \bibinfo
  {pages} {044907} (\bibinfo {year} {2024})},\ \Eprint
  {http://arxiv.org/abs/2301.02485} {arXiv:2301.02485 [hep-ph]} \BibitemShut
  {NoStop}%
\bibitem [{\citenamefont {Milhano}\ \emph {et~al.}(2018)\citenamefont
  {Milhano}, \citenamefont {Wiedemann},\ and\ \citenamefont
  {Zapp}}]{Milhano:2017nzm}%
  \BibitemOpen
  \bibfield  {author} {\bibinfo {author} {\bibfnamefont {G.}~\bibnamefont
  {Milhano}}, \bibinfo {author} {\bibfnamefont {U.~A.}\ \bibnamefont
  {Wiedemann}}, \ and\ \bibinfo {author} {\bibfnamefont {K.~C.}\ \bibnamefont
  {Zapp}},\ }\href {\doibase 10.1016/j.physletb.2018.01.029} {\bibfield
  {journal} {\bibinfo  {journal} {Phys. Lett. B}\ }\textbf {\bibinfo {volume}
  {779}},\ \bibinfo {pages} {409} (\bibinfo {year} {2018})},\ \Eprint
  {http://arxiv.org/abs/1707.04142} {arXiv:1707.04142 [hep-ph]} \BibitemShut
  {NoStop}%
\bibitem [{\citenamefont {Caucal}\ \emph {et~al.}(2019)\citenamefont {Caucal},
  \citenamefont {Iancu},\ and\ \citenamefont {Soyez}}]{Caucal:2019uvr}%
  \BibitemOpen
  \bibfield  {author} {\bibinfo {author} {\bibfnamefont {P.}~\bibnamefont
  {Caucal}}, \bibinfo {author} {\bibfnamefont {E.}~\bibnamefont {Iancu}}, \
  and\ \bibinfo {author} {\bibfnamefont {G.}~\bibnamefont {Soyez}},\ }\href
  {\doibase 10.1007/JHEP10(2019)273} {\bibfield  {journal} {\bibinfo  {journal}
  {JHEP}\ }\textbf {\bibinfo {volume} {10}},\ \bibinfo {pages} {273} (\bibinfo
  {year} {2019})},\ \Eprint {http://arxiv.org/abs/1907.04866} {arXiv:1907.04866
  [hep-ph]} \BibitemShut {NoStop}%
\bibitem [{\citenamefont {Casalderrey-Solana}\ \emph
  {et~al.}(2020)\citenamefont {Casalderrey-Solana}, \citenamefont {Milhano},
  \citenamefont {Pablos},\ and\ \citenamefont
  {Rajagopal}}]{Casalderrey-Solana:2019ubu}%
  \BibitemOpen
  \bibfield  {author} {\bibinfo {author} {\bibfnamefont {J.}~\bibnamefont
  {Casalderrey-Solana}}, \bibinfo {author} {\bibfnamefont {G.}~\bibnamefont
  {Milhano}}, \bibinfo {author} {\bibfnamefont {D.}~\bibnamefont {Pablos}}, \
  and\ \bibinfo {author} {\bibfnamefont {K.}~\bibnamefont {Rajagopal}},\ }\href
  {\doibase 10.1007/JHEP01(2020)044} {\bibfield  {journal} {\bibinfo  {journal}
  {JHEP}\ }\textbf {\bibinfo {volume} {01}},\ \bibinfo {pages} {044} (\bibinfo
  {year} {2020})},\ \Eprint {http://arxiv.org/abs/1907.11248} {arXiv:1907.11248
  [hep-ph]} \BibitemShut {NoStop}%
\bibitem [{\citenamefont {Wang}\ \emph {et~al.}(2019)\citenamefont {Wang},
  \citenamefont {Dai}, \citenamefont {Zhang},\ and\ \citenamefont
  {Wang}}]{Wang:2019xey}%
  \BibitemOpen
  \bibfield  {author} {\bibinfo {author} {\bibfnamefont {S.}~\bibnamefont
  {Wang}}, \bibinfo {author} {\bibfnamefont {W.}~\bibnamefont {Dai}}, \bibinfo
  {author} {\bibfnamefont {B.-W.}\ \bibnamefont {Zhang}}, \ and\ \bibinfo
  {author} {\bibfnamefont {E.}~\bibnamefont {Wang}},\ }\href {\doibase
  10.1140/epjc/s10052-019-7312-4} {\bibfield  {journal} {\bibinfo  {journal}
  {Eur. Phys. J. C}\ }\textbf {\bibinfo {volume} {79}},\ \bibinfo {pages} {789}
  (\bibinfo {year} {2019})},\ \Eprint {http://arxiv.org/abs/1906.01499}
  {arXiv:1906.01499 [nucl-th]} \BibitemShut {NoStop}%
\bibitem [{\citenamefont {Li}\ \emph {et~al.}(2023)\citenamefont {Li},
  \citenamefont {Wang},\ and\ \citenamefont {Zhang}}]{Li:2022tcr}%
  \BibitemOpen
  \bibfield  {author} {\bibinfo {author} {\bibfnamefont {Y.}~\bibnamefont
  {Li}}, \bibinfo {author} {\bibfnamefont {S.}~\bibnamefont {Wang}}, \ and\
  \bibinfo {author} {\bibfnamefont {B.-W.}\ \bibnamefont {Zhang}},\ }\href
  {\doibase 10.1103/PhysRevC.108.024905} {\bibfield  {journal} {\bibinfo
  {journal} {Phys. Rev. C}\ }\textbf {\bibinfo {volume} {108}},\ \bibinfo
  {pages} {024905} (\bibinfo {year} {2023})},\ \Eprint
  {http://arxiv.org/abs/2209.00548} {arXiv:2209.00548 [hep-ph]} \BibitemShut
  {NoStop}%
\bibitem [{\citenamefont {Budhraja}\ \emph {et~al.}(2025)\citenamefont
  {Budhraja}, \citenamefont {Sharma},\ and\ \citenamefont
  {Singh}}]{Budhraja:2023rgo}%
  \BibitemOpen
  \bibfield  {author} {\bibinfo {author} {\bibfnamefont {A.}~\bibnamefont
  {Budhraja}}, \bibinfo {author} {\bibfnamefont {R.}~\bibnamefont {Sharma}}, \
  and\ \bibinfo {author} {\bibfnamefont {B.}~\bibnamefont {Singh}},\ }\href
  {\doibase 10.1103/r12p-3jbm} {\bibfield  {journal} {\bibinfo  {journal}
  {Phys. Rev. D}\ }\textbf {\bibinfo {volume} {112}},\ \bibinfo {pages}
  {034017} (\bibinfo {year} {2025})},\ \Eprint
  {http://arxiv.org/abs/2305.10237} {arXiv:2305.10237 [hep-ph]} \BibitemShut
  {NoStop}%
\bibitem [{\citenamefont {Abdallah}\ \emph {et~al.}(2022)\citenamefont
  {Abdallah} \emph {et~al.}}]{STAR:2021kjt}%
  \BibitemOpen
  \bibfield  {author} {\bibinfo {author} {\bibfnamefont {M.~S.}\ \bibnamefont
  {Abdallah}} \emph {et~al.} (\bibinfo {collaboration} {STAR}),\ }\href
  {\doibase 10.1103/PhysRevC.105.044906} {\bibfield  {journal} {\bibinfo
  {journal} {Phys. Rev. C}\ }\textbf {\bibinfo {volume} {105}},\ \bibinfo
  {pages} {044906} (\bibinfo {year} {2022})},\ \Eprint
  {http://arxiv.org/abs/2109.09793} {arXiv:2109.09793 [nucl-ex]} \BibitemShut
  {NoStop}%
\bibitem [{\citenamefont {Acharya}\ \emph {et~al.}(2022)\citenamefont {Acharya}
  \emph {et~al.}}]{ALargeIonColliderExperiment:2021mqf}%
  \BibitemOpen
  \bibfield  {author} {\bibinfo {author} {\bibfnamefont {S.}~\bibnamefont
  {Acharya}} \emph {et~al.} (\bibinfo {collaboration} {A Large Ion Collider
  Experiment, ALICE}),\ }\href {\doibase 10.1103/PhysRevLett.128.102001}
  {\bibfield  {journal} {\bibinfo  {journal} {Phys. Rev. Lett.}\ }\textbf
  {\bibinfo {volume} {128}},\ \bibinfo {pages} {102001} (\bibinfo {year}
  {2022})},\ \Eprint {http://arxiv.org/abs/2107.12984} {arXiv:2107.12984
  [nucl-ex]} \BibitemShut {NoStop}%
\bibitem [{\citenamefont {Aad}\ \emph {et~al.}(2023{\natexlab{a}})\citenamefont
  {Aad} \emph {et~al.}}]{ATLAS:2022vii}%
  \BibitemOpen
  \bibfield  {author} {\bibinfo {author} {\bibfnamefont {G.}~\bibnamefont
  {Aad}} \emph {et~al.} (\bibinfo {collaboration} {ATLAS}),\ }\href {\doibase
  10.1103/PhysRevC.107.054909} {\bibfield  {journal} {\bibinfo  {journal}
  {Phys. Rev. C}\ }\textbf {\bibinfo {volume} {107}},\ \bibinfo {pages}
  {054909} (\bibinfo {year} {2023}{\natexlab{a}})},\ \Eprint
  {http://arxiv.org/abs/2211.11470} {arXiv:2211.11470 [nucl-ex]} \BibitemShut
  {NoStop}%
\bibitem [{\citenamefont {Acharya}\ \emph {et~al.}(2018)\citenamefont {Acharya}
  \emph {et~al.}}]{ALICE:2018dxf}%
  \BibitemOpen
  \bibfield  {author} {\bibinfo {author} {\bibfnamefont {S.}~\bibnamefont
  {Acharya}} \emph {et~al.} (\bibinfo {collaboration} {ALICE}),\ }\href
  {\doibase 10.1007/JHEP10(2018)139} {\bibfield  {journal} {\bibinfo  {journal}
  {JHEP}\ }\textbf {\bibinfo {volume} {10}},\ \bibinfo {pages} {139} (\bibinfo
  {year} {2018})},\ \Eprint {http://arxiv.org/abs/1807.06854} {arXiv:1807.06854
  [nucl-ex]} \BibitemShut {NoStop}%
\bibitem [{\citenamefont {Acharya}\ \emph {et~al.}(2023)\citenamefont {Acharya}
  \emph {et~al.}}]{ALICE:2023dwg}%
  \BibitemOpen
  \bibfield  {author} {\bibinfo {author} {\bibfnamefont {S.}~\bibnamefont
  {Acharya}} \emph {et~al.} (\bibinfo {collaboration} {ALICE}),\ }\href@noop {}
  {\  (\bibinfo {year} {2023})},\ \Eprint {http://arxiv.org/abs/2303.13347}
  {arXiv:2303.13347 [nucl-ex]} \BibitemShut {NoStop}%
\bibitem [{\citenamefont {Ehlers}(2022)}]{Ehlers:2022dfp}%
  \BibitemOpen
  \bibfield  {author} {\bibinfo {author} {\bibfnamefont {R.}~\bibnamefont
  {Ehlers}} (\bibinfo {collaboration} {ALICE}),\ }\href {\doibase
  10.22323/1.414.0460} {\bibfield  {journal} {\bibinfo  {journal} {PoS}\
  }\textbf {\bibinfo {volume} {ICHEP2022}},\ \bibinfo {pages} {460} (\bibinfo
  {year} {2022})},\ \Eprint {http://arxiv.org/abs/2211.11800} {arXiv:2211.11800
  [nucl-ex]} \BibitemShut {NoStop}%
\bibitem [{\citenamefont {Aad}\ \emph {et~al.}(2023{\natexlab{b}})\citenamefont
  {Aad} \emph {et~al.}}]{ATLAS:2023hso}%
  \BibitemOpen
  \bibfield  {author} {\bibinfo {author} {\bibfnamefont {G.}~\bibnamefont
  {Aad}} \emph {et~al.} (\bibinfo {collaboration} {ATLAS}),\ }\href {\doibase
  10.1103/PhysRevLett.131.172301} {\bibfield  {journal} {\bibinfo  {journal}
  {Phys. Rev. Lett.}\ }\textbf {\bibinfo {volume} {131}},\ \bibinfo {pages}
  {172301} (\bibinfo {year} {2023}{\natexlab{b}})},\ \Eprint
  {http://arxiv.org/abs/2301.05606} {arXiv:2301.05606 [nucl-ex]} \BibitemShut
  {NoStop}%
\bibitem [{\citenamefont {Baier}\ \emph {et~al.}(2001)\citenamefont {Baier},
  \citenamefont {Dokshitzer}, \citenamefont {Mueller},\ and\ \citenamefont
  {Schiff}}]{Baier:2001yt}%
  \BibitemOpen
  \bibfield  {author} {\bibinfo {author} {\bibfnamefont {R.}~\bibnamefont
  {Baier}}, \bibinfo {author} {\bibfnamefont {Y.~L.}\ \bibnamefont
  {Dokshitzer}}, \bibinfo {author} {\bibfnamefont {A.~H.}\ \bibnamefont
  {Mueller}}, \ and\ \bibinfo {author} {\bibfnamefont {D.}~\bibnamefont
  {Schiff}},\ }\href {\doibase 10.1088/1126-6708/2001/09/033} {\bibfield
  {journal} {\bibinfo  {journal} {JHEP}\ }\textbf {\bibinfo {volume} {09}},\
  \bibinfo {pages} {033} (\bibinfo {year} {2001})},\ \Eprint
  {http://arxiv.org/abs/hep-ph/0106347} {arXiv:hep-ph/0106347} \BibitemShut
  {NoStop}%
\bibitem [{\citenamefont {Renk}(2013)}]{Renk:2012ve}%
  \BibitemOpen
  \bibfield  {author} {\bibinfo {author} {\bibfnamefont {T.}~\bibnamefont
  {Renk}},\ }\href {\doibase 10.1103/PhysRevC.88.054902} {\bibfield  {journal}
  {\bibinfo  {journal} {Phys. Rev. C}\ }\textbf {\bibinfo {volume} {88}},\
  \bibinfo {pages} {054902} (\bibinfo {year} {2013})},\ \Eprint
  {http://arxiv.org/abs/1212.0646} {arXiv:1212.0646 [hep-ph]} \BibitemShut
  {NoStop}%
\bibitem [{\citenamefont {Wang}\ \emph {et~al.}(2021)\citenamefont {Wang},
  \citenamefont {Kang}, \citenamefont {Dai}, \citenamefont {Zhang},\ and\
  \citenamefont {Wang}}]{Wang:2021jgm}%
  \BibitemOpen
  \bibfield  {author} {\bibinfo {author} {\bibfnamefont {S.}~\bibnamefont
  {Wang}}, \bibinfo {author} {\bibfnamefont {J.-W.}\ \bibnamefont {Kang}},
  \bibinfo {author} {\bibfnamefont {W.}~\bibnamefont {Dai}}, \bibinfo {author}
  {\bibfnamefont {B.-W.}\ \bibnamefont {Zhang}}, \ and\ \bibinfo {author}
  {\bibfnamefont {E.}~\bibnamefont {Wang}},\ }\href {\doibase
  10.1140/epja/s10050-022-00785-9} {\bibfield  {journal} {\bibinfo  {journal}
  {Eur. Phys. J. A}\ }\textbf {\bibinfo {volume} {58}},\ \bibinfo {pages} {149}
  (\bibinfo {year} {2021})},\ \Eprint {http://arxiv.org/abs/2107.12000}
  {arXiv:2107.12000 [nucl-th]} \BibitemShut {NoStop}%
\bibitem [{\citenamefont {Brewer}\ \emph {et~al.}(2022)\citenamefont {Brewer},
  \citenamefont {Brodsky},\ and\ \citenamefont {Rajagopal}}]{Brewer:2021hmh}%
  \BibitemOpen
  \bibfield  {author} {\bibinfo {author} {\bibfnamefont {J.}~\bibnamefont
  {Brewer}}, \bibinfo {author} {\bibfnamefont {Q.}~\bibnamefont {Brodsky}}, \
  and\ \bibinfo {author} {\bibfnamefont {K.}~\bibnamefont {Rajagopal}},\ }\href
  {\doibase 10.1007/JHEP02(2022)175} {\bibfield  {journal} {\bibinfo  {journal}
  {JHEP}\ }\textbf {\bibinfo {volume} {02}},\ \bibinfo {pages} {175} (\bibinfo
  {year} {2022})},\ \Eprint {http://arxiv.org/abs/2110.13159} {arXiv:2110.13159
  [hep-ph]} \BibitemShut {NoStop}%
\bibitem [{\citenamefont {Zhang}\ \emph
  {et~al.}(2022{\natexlab{b}})\citenamefont {Zhang}, \citenamefont {Yang},\
  and\ \citenamefont {Zhang}}]{Zhang:2021sua}%
  \BibitemOpen
  \bibfield  {author} {\bibinfo {author} {\bibfnamefont {S.-L.}\ \bibnamefont
  {Zhang}}, \bibinfo {author} {\bibfnamefont {M.-Q.}\ \bibnamefont {Yang}}, \
  and\ \bibinfo {author} {\bibfnamefont {B.-W.}\ \bibnamefont {Zhang}},\ }\href
  {\doibase 10.1140/epjc/s10052-022-10340-x} {\bibfield  {journal} {\bibinfo
  {journal} {Eur. Phys. J. C}\ }\textbf {\bibinfo {volume} {82}},\ \bibinfo
  {pages} {414} (\bibinfo {year} {2022}{\natexlab{b}})},\ \Eprint
  {http://arxiv.org/abs/2105.04955} {arXiv:2105.04955 [hep-ph]} \BibitemShut
  {NoStop}%
\bibitem [{\citenamefont {Brewer}\ \emph {et~al.}(2019)\citenamefont {Brewer},
  \citenamefont {Milhano},\ and\ \citenamefont {Thaler}}]{Brewer:2018dfs}%
  \BibitemOpen
  \bibfield  {author} {\bibinfo {author} {\bibfnamefont {J.}~\bibnamefont
  {Brewer}}, \bibinfo {author} {\bibfnamefont {J.~G.}\ \bibnamefont {Milhano}},
  \ and\ \bibinfo {author} {\bibfnamefont {J.}~\bibnamefont {Thaler}},\ }\href
  {\doibase 10.1103/PhysRevLett.122.222301} {\bibfield  {journal} {\bibinfo
  {journal} {Phys. Rev. Lett.}\ }\textbf {\bibinfo {volume} {122}},\ \bibinfo
  {pages} {222301} (\bibinfo {year} {2019})},\ \Eprint
  {http://arxiv.org/abs/1812.05111} {arXiv:1812.05111 [hep-ph]} \BibitemShut
  {NoStop}%
\bibitem [{\citenamefont {Du}\ \emph {et~al.}(2020)\citenamefont {Du},
  \citenamefont {Pablos},\ and\ \citenamefont {Tywoniuk}}]{Du:2020pmp}%
  \BibitemOpen
  \bibfield  {author} {\bibinfo {author} {\bibfnamefont {Y.-L.}\ \bibnamefont
  {Du}}, \bibinfo {author} {\bibfnamefont {D.}~\bibnamefont {Pablos}}, \ and\
  \bibinfo {author} {\bibfnamefont {K.}~\bibnamefont {Tywoniuk}},\ }\href
  {\doibase 10.1007/JHEP03(2021)206} {\bibfield  {journal} {\bibinfo  {journal}
  {JHEP}\ }\textbf {\bibinfo {volume} {21}},\ \bibinfo {pages} {206} (\bibinfo
  {year} {2020})},\ \Eprint {http://arxiv.org/abs/2012.07797} {arXiv:2012.07797
  [hep-ph]} \BibitemShut {NoStop}%
\bibitem [{\citenamefont {Neufeld}\ \emph {et~al.}(2011)\citenamefont
  {Neufeld}, \citenamefont {Vitev},\ and\ \citenamefont
  {Zhang}}]{Neufeld:2010fj}%
  \BibitemOpen
  \bibfield  {author} {\bibinfo {author} {\bibfnamefont {R.~B.}\ \bibnamefont
  {Neufeld}}, \bibinfo {author} {\bibfnamefont {I.}~\bibnamefont {Vitev}}, \
  and\ \bibinfo {author} {\bibfnamefont {B.~W.}\ \bibnamefont {Zhang}},\ }\href
  {\doibase 10.1103/PhysRevC.83.034902} {\bibfield  {journal} {\bibinfo
  {journal} {Phys. Rev. C}\ }\textbf {\bibinfo {volume} {83}},\ \bibinfo
  {pages} {034902} (\bibinfo {year} {2011})},\ \Eprint
  {http://arxiv.org/abs/1006.2389} {arXiv:1006.2389 [hep-ph]} \BibitemShut
  {NoStop}%
\bibitem [{\citenamefont {Wang}\ \emph {et~al.}(1996)\citenamefont {Wang},
  \citenamefont {Huang},\ and\ \citenamefont {Sarcevic}}]{Wang:1996yh}%
  \BibitemOpen
  \bibfield  {author} {\bibinfo {author} {\bibfnamefont {X.-N.}\ \bibnamefont
  {Wang}}, \bibinfo {author} {\bibfnamefont {Z.}~\bibnamefont {Huang}}, \ and\
  \bibinfo {author} {\bibfnamefont {I.}~\bibnamefont {Sarcevic}},\ }\href
  {\doibase 10.1103/PhysRevLett.77.231} {\bibfield  {journal} {\bibinfo
  {journal} {Phys. Rev. Lett.}\ }\textbf {\bibinfo {volume} {77}},\ \bibinfo
  {pages} {231} (\bibinfo {year} {1996})},\ \Eprint
  {http://arxiv.org/abs/hep-ph/9605213} {arXiv:hep-ph/9605213} \BibitemShut
  {NoStop}%
\bibitem [{\citenamefont {Dai}\ \emph {et~al.}(2013)\citenamefont {Dai},
  \citenamefont {Vitev},\ and\ \citenamefont {Zhang}}]{Dai:2012am}%
  \BibitemOpen
  \bibfield  {author} {\bibinfo {author} {\bibfnamefont {W.}~\bibnamefont
  {Dai}}, \bibinfo {author} {\bibfnamefont {I.}~\bibnamefont {Vitev}}, \ and\
  \bibinfo {author} {\bibfnamefont {B.-W.}\ \bibnamefont {Zhang}},\ }\href
  {\doibase 10.1103/PhysRevLett.110.142001} {\bibfield  {journal} {\bibinfo
  {journal} {Phys. Rev. Lett.}\ }\textbf {\bibinfo {volume} {110}},\ \bibinfo
  {pages} {142001} (\bibinfo {year} {2013})},\ \Eprint
  {http://arxiv.org/abs/1207.5177} {arXiv:1207.5177 [hep-ph]} \BibitemShut
  {NoStop}%
\bibitem [{\citenamefont {Chen}\ \emph {et~al.}(2018)\citenamefont {Chen},
  \citenamefont {Cao}, \citenamefont {Luo}, \citenamefont {Pang},\ and\
  \citenamefont {Wang}}]{Chen:2017zte}%
  \BibitemOpen
  \bibfield  {author} {\bibinfo {author} {\bibfnamefont {W.}~\bibnamefont
  {Chen}}, \bibinfo {author} {\bibfnamefont {S.}~\bibnamefont {Cao}}, \bibinfo
  {author} {\bibfnamefont {T.}~\bibnamefont {Luo}}, \bibinfo {author}
  {\bibfnamefont {L.-G.}\ \bibnamefont {Pang}}, \ and\ \bibinfo {author}
  {\bibfnamefont {X.-N.}\ \bibnamefont {Wang}},\ }\href {\doibase
  10.1016/j.physletb.2017.12.015} {\bibfield  {journal} {\bibinfo  {journal}
  {Phys. Lett. B}\ }\textbf {\bibinfo {volume} {777}},\ \bibinfo {pages} {86}
  (\bibinfo {year} {2018})},\ \Eprint {http://arxiv.org/abs/1704.03648}
  {arXiv:1704.03648 [nucl-th]} \BibitemShut {NoStop}%
\bibitem [{\citenamefont {Wang}\ and\ \citenamefont
  {Zhu}(2013)}]{Wang:2013cia}%
  \BibitemOpen
  \bibfield  {author} {\bibinfo {author} {\bibfnamefont {X.-N.}\ \bibnamefont
  {Wang}}\ and\ \bibinfo {author} {\bibfnamefont {Y.}~\bibnamefont {Zhu}},\
  }\href {\doibase 10.1103/PhysRevLett.111.062301} {\bibfield  {journal}
  {\bibinfo  {journal} {Phys. Rev. Lett.}\ }\textbf {\bibinfo {volume} {111}},\
  \bibinfo {pages} {062301} (\bibinfo {year} {2013})},\ \Eprint
  {http://arxiv.org/abs/1302.5874} {arXiv:1302.5874 [hep-ph]} \BibitemShut
  {NoStop}%
\bibitem [{\citenamefont {Zhang}\ \emph {et~al.}(2018)\citenamefont {Zhang},
  \citenamefont {Luo}, \citenamefont {Wang},\ and\ \citenamefont
  {Zhang}}]{Zhang:2018urd}%
  \BibitemOpen
  \bibfield  {author} {\bibinfo {author} {\bibfnamefont {S.-L.}\ \bibnamefont
  {Zhang}}, \bibinfo {author} {\bibfnamefont {T.}~\bibnamefont {Luo}}, \bibinfo
  {author} {\bibfnamefont {X.-N.}\ \bibnamefont {Wang}}, \ and\ \bibinfo
  {author} {\bibfnamefont {B.-W.}\ \bibnamefont {Zhang}},\ }\href {\doibase
  10.1103/PhysRevC.98.021901} {\bibfield  {journal} {\bibinfo  {journal} {Phys.
  Rev. C}\ }\textbf {\bibinfo {volume} {98}},\ \bibinfo {pages} {021901}
  (\bibinfo {year} {2018})},\ \Eprint {http://arxiv.org/abs/1804.11041}
  {arXiv:1804.11041 [nucl-th]} \BibitemShut {NoStop}%
\bibitem [{\citenamefont {Chang}\ \emph {et~al.}(2020)\citenamefont {Chang},
  \citenamefont {Tachibana},\ and\ \citenamefont {Qin}}]{Chang:2019sae}%
  \BibitemOpen
  \bibfield  {author} {\bibinfo {author} {\bibfnamefont {N.-B.}\ \bibnamefont
  {Chang}}, \bibinfo {author} {\bibfnamefont {Y.}~\bibnamefont {Tachibana}}, \
  and\ \bibinfo {author} {\bibfnamefont {G.-Y.}\ \bibnamefont {Qin}},\ }\href
  {\doibase 10.1016/j.physletb.2019.135181} {\bibfield  {journal} {\bibinfo
  {journal} {Phys. Lett. B}\ }\textbf {\bibinfo {volume} {801}},\ \bibinfo
  {pages} {135181} (\bibinfo {year} {2020})},\ \Eprint
  {http://arxiv.org/abs/1906.09562} {arXiv:1906.09562 [nucl-th]} \BibitemShut
  {NoStop}%
\bibitem [{\citenamefont {Sirimanna}\ \emph {et~al.}(2024)\citenamefont
  {Sirimanna} \emph {et~al.}}]{JETSCAPE:2024rma}%
  \BibitemOpen
  \bibfield  {author} {\bibinfo {author} {\bibfnamefont {C.}~\bibnamefont
  {Sirimanna}} \emph {et~al.} (\bibinfo {collaboration} {JETSCAPE}),\ }\href
  {\doibase 10.1051/epjconf/202429611008} {\bibfield  {journal} {\bibinfo
  {journal} {EPJ Web Conf.}\ }\textbf {\bibinfo {volume} {296}},\ \bibinfo
  {pages} {11008} (\bibinfo {year} {2024})},\ \Eprint
  {http://arxiv.org/abs/2401.17259} {arXiv:2401.17259 [nucl-th]} \BibitemShut
  {NoStop}%
\bibitem [{\citenamefont {Wang}\ \emph {et~al.}(2023)\citenamefont {Wang},
  \citenamefont {Dai}, \citenamefont {Zhang},\ and\ \citenamefont
  {Wang}}]{Wang:2020qwe}%
  \BibitemOpen
  \bibfield  {author} {\bibinfo {author} {\bibfnamefont {S.}~\bibnamefont
  {Wang}}, \bibinfo {author} {\bibfnamefont {W.}~\bibnamefont {Dai}}, \bibinfo
  {author} {\bibfnamefont {B.-W.}\ \bibnamefont {Zhang}}, \ and\ \bibinfo
  {author} {\bibfnamefont {E.}~\bibnamefont {Wang}},\ }\href {\doibase
  10.1088/1674-1137/acc1ca} {\bibfield  {journal} {\bibinfo  {journal} {Chin.
  Phys. C}\ }\textbf {\bibinfo {volume} {47}},\ \bibinfo {pages} {054102}
  (\bibinfo {year} {2023})},\ \Eprint {http://arxiv.org/abs/2005.07018}
  {arXiv:2005.07018 [hep-ph]} \BibitemShut {NoStop}%
\bibitem [{\citenamefont {Aboona}\ \emph
  {et~al.}(2025{\natexlab{a}})\citenamefont {Aboona} \emph
  {et~al.}}]{STAR:2023pal}%
  \BibitemOpen
  \bibfield  {author} {\bibinfo {author} {\bibfnamefont {B.~E.}\ \bibnamefont
  {Aboona}} \emph {et~al.} (\bibinfo {collaboration} {STAR}),\ }\href {\doibase
  10.1103/PhysRevLett.134.232301} {\bibfield  {journal} {\bibinfo  {journal}
  {Phys. Rev. Lett.}\ }\textbf {\bibinfo {volume} {134}},\ \bibinfo {pages}
  {232301} (\bibinfo {year} {2025}{\natexlab{a}})},\ \Eprint
  {http://arxiv.org/abs/2309.00156} {arXiv:2309.00156 [nucl-ex]} \BibitemShut
  {NoStop}%
\bibitem [{\citenamefont {Aboona}\ \emph
  {et~al.}(2025{\natexlab{b}})\citenamefont {Aboona} \emph
  {et~al.}}]{STAR:2023ksv}%
  \BibitemOpen
  \bibfield  {author} {\bibinfo {author} {\bibfnamefont {B.~E.}\ \bibnamefont
  {Aboona}} \emph {et~al.} (\bibinfo {collaboration} {STAR}),\ }\href {\doibase
  10.1103/8b8y-98yh} {\bibfield  {journal} {\bibinfo  {journal} {Phys. Rev. C}\
  }\textbf {\bibinfo {volume} {111}},\ \bibinfo {pages} {064907} (\bibinfo
  {year} {2025}{\natexlab{b}})},\ \Eprint {http://arxiv.org/abs/2309.00145}
  {arXiv:2309.00145 [nucl-ex]} \BibitemShut {NoStop}%
\bibitem [{\citenamefont {Larkoski}\ \emph
  {et~al.}(2014{\natexlab{a}})\citenamefont {Larkoski}, \citenamefont
  {Thaler},\ and\ \citenamefont {Waalewijn}}]{Larkoski:2014pca}%
  \BibitemOpen
  \bibfield  {author} {\bibinfo {author} {\bibfnamefont {A.~J.}\ \bibnamefont
  {Larkoski}}, \bibinfo {author} {\bibfnamefont {J.}~\bibnamefont {Thaler}}, \
  and\ \bibinfo {author} {\bibfnamefont {W.~J.}\ \bibnamefont {Waalewijn}},\
  }\href {\doibase 10.1007/JHEP11(2014)129} {\bibfield  {journal} {\bibinfo
  {journal} {JHEP}\ }\textbf {\bibinfo {volume} {11}},\ \bibinfo {pages} {129}
  (\bibinfo {year} {2014}{\natexlab{a}})},\ \Eprint
  {http://arxiv.org/abs/1408.3122} {arXiv:1408.3122 [hep-ph]} \BibitemShut
  {NoStop}%
\bibitem [{\citenamefont {Larkoski}\ \emph
  {et~al.}(2014{\natexlab{b}})\citenamefont {Larkoski}, \citenamefont
  {Marzani}, \citenamefont {Soyez},\ and\ \citenamefont
  {Thaler}}]{Larkoski:2014wba}%
  \BibitemOpen
  \bibfield  {author} {\bibinfo {author} {\bibfnamefont {A.~J.}\ \bibnamefont
  {Larkoski}}, \bibinfo {author} {\bibfnamefont {S.}~\bibnamefont {Marzani}},
  \bibinfo {author} {\bibfnamefont {G.}~\bibnamefont {Soyez}}, \ and\ \bibinfo
  {author} {\bibfnamefont {J.}~\bibnamefont {Thaler}},\ }\href {\doibase
  10.1007/JHEP05(2014)146} {\bibfield  {journal} {\bibinfo  {journal} {JHEP}\
  }\textbf {\bibinfo {volume} {05}},\ \bibinfo {pages} {146} (\bibinfo {year}
  {2014}{\natexlab{b}})},\ \Eprint {http://arxiv.org/abs/1402.2657}
  {arXiv:1402.2657 [hep-ph]} \BibitemShut {NoStop}%
\bibitem [{\citenamefont {Hayrapetyan}\ \emph
  {et~al.}(2025{\natexlab{b}})\citenamefont {Hayrapetyan} \emph
  {et~al.}}]{CMS:2024zjn}%
  \BibitemOpen
  \bibfield  {author} {\bibinfo {author} {\bibfnamefont {A.}~\bibnamefont
  {Hayrapetyan}} \emph {et~al.} (\bibinfo {collaboration} {CMS}),\ }\href
  {\doibase 10.1016/j.physletb.2024.139088} {\bibfield  {journal} {\bibinfo
  {journal} {Phys. Lett. B}\ }\textbf {\bibinfo {volume} {861}},\ \bibinfo
  {pages} {139088} (\bibinfo {year} {2025}{\natexlab{b}})},\ \Eprint
  {http://arxiv.org/abs/2405.02737} {arXiv:2405.02737 [nucl-ex]} \BibitemShut
  {NoStop}%
\bibitem [{\citenamefont {Sj{\"o}strand}\ \emph {et~al.}(2015)\citenamefont
  {Sj{\"o}strand}, \citenamefont {Ask}, \citenamefont {Christiansen},
  \citenamefont {Corke}, \citenamefont {Desai}, \citenamefont {Ilten},
  \citenamefont {Mrenna}, \citenamefont {Prestel}, \citenamefont {Rasmussen},\
  and\ \citenamefont {Skands}}]{Sjostrand:2014zea}%
  \BibitemOpen
  \bibfield  {author} {\bibinfo {author} {\bibfnamefont {T.}~\bibnamefont
  {Sj{\"o}strand}}, \bibinfo {author} {\bibfnamefont {S.}~\bibnamefont {Ask}},
  \bibinfo {author} {\bibfnamefont {J.~R.}\ \bibnamefont {Christiansen}},
  \bibinfo {author} {\bibfnamefont {R.}~\bibnamefont {Corke}}, \bibinfo
  {author} {\bibfnamefont {N.}~\bibnamefont {Desai}}, \bibinfo {author}
  {\bibfnamefont {P.}~\bibnamefont {Ilten}}, \bibinfo {author} {\bibfnamefont
  {S.}~\bibnamefont {Mrenna}}, \bibinfo {author} {\bibfnamefont
  {S.}~\bibnamefont {Prestel}}, \bibinfo {author} {\bibfnamefont {C.~O.}\
  \bibnamefont {Rasmussen}}, \ and\ \bibinfo {author} {\bibfnamefont {P.~Z.}\
  \bibnamefont {Skands}},\ }\href {\doibase 10.1016/j.cpc.2015.01.024}
  {\bibfield  {journal} {\bibinfo  {journal} {Comput. Phys. Commun.}\ }\textbf
  {\bibinfo {volume} {191}},\ \bibinfo {pages} {159} (\bibinfo {year}
  {2015})},\ \Eprint {http://arxiv.org/abs/1410.3012} {arXiv:1410.3012
  [hep-ph]} \BibitemShut {NoStop}%
\bibitem [{\citenamefont {Skands}\ \emph {et~al.}(2014)\citenamefont {Skands},
  \citenamefont {Carrazza},\ and\ \citenamefont {Rojo}}]{Skands:2014pea}%
  \BibitemOpen
  \bibfield  {author} {\bibinfo {author} {\bibfnamefont {P.}~\bibnamefont
  {Skands}}, \bibinfo {author} {\bibfnamefont {S.}~\bibnamefont {Carrazza}}, \
  and\ \bibinfo {author} {\bibfnamefont {J.}~\bibnamefont {Rojo}},\ }\href
  {\doibase 10.1140/epjc/s10052-014-3024-y} {\bibfield  {journal} {\bibinfo
  {journal} {Eur. Phys. J. C}\ }\textbf {\bibinfo {volume} {74}},\ \bibinfo
  {pages} {3024} (\bibinfo {year} {2014})},\ \Eprint
  {http://arxiv.org/abs/1404.5630} {arXiv:1404.5630 [hep-ph]} \BibitemShut
  {NoStop}%
\bibitem [{\citenamefont {Dai}\ \emph {et~al.}(2020)\citenamefont {Dai},
  \citenamefont {Wang}, \citenamefont {Zhang}, \citenamefont {Zhang},\ and\
  \citenamefont {Wang}}]{Dai:2018mhw}%
  \BibitemOpen
  \bibfield  {author} {\bibinfo {author} {\bibfnamefont {W.}~\bibnamefont
  {Dai}}, \bibinfo {author} {\bibfnamefont {S.}~\bibnamefont {Wang}}, \bibinfo
  {author} {\bibfnamefont {S.-L.}\ \bibnamefont {Zhang}}, \bibinfo {author}
  {\bibfnamefont {B.-W.}\ \bibnamefont {Zhang}}, \ and\ \bibinfo {author}
  {\bibfnamefont {E.}~\bibnamefont {Wang}},\ }\href {\doibase
  10.1088/1674-1137/abab8f} {\bibfield  {journal} {\bibinfo  {journal} {Chin.
  Phys. C}\ }\textbf {\bibinfo {volume} {44}},\ \bibinfo {pages} {104105}
  (\bibinfo {year} {2020})},\ \Eprint {http://arxiv.org/abs/1806.06332}
  {arXiv:1806.06332 [nucl-th]} \BibitemShut {NoStop}%
\bibitem [{\citenamefont {Li}\ \emph {et~al.}(2024)\citenamefont {Li},
  \citenamefont {Shen}, \citenamefont {Wang},\ and\ \citenamefont
  {Zhang}}]{Li:2024uzk}%
  \BibitemOpen
  \bibfield  {author} {\bibinfo {author} {\bibfnamefont {Y.}~\bibnamefont
  {Li}}, \bibinfo {author} {\bibfnamefont {S.}~\bibnamefont {Shen}}, \bibinfo
  {author} {\bibfnamefont {S.}~\bibnamefont {Wang}}, \ and\ \bibinfo {author}
  {\bibfnamefont {B.-W.}\ \bibnamefont {Zhang}},\ }\href {\doibase
  10.1007/s41365-024-01482-6} {\bibfield  {journal} {\bibinfo  {journal} {Nucl.
  Sci. Tech.}\ }\textbf {\bibinfo {volume} {35}},\ \bibinfo {pages} {113}
  (\bibinfo {year} {2024})},\ \Eprint {http://arxiv.org/abs/2401.01706}
  {arXiv:2401.01706 [hep-ph]} \BibitemShut {NoStop}%
\bibitem [{\citenamefont {Wang}\ \emph
  {et~al.}(2025{\natexlab{a}})\citenamefont {Wang}, \citenamefont {Li},
  \citenamefont {Li}, \citenamefont {Zhang},\ and\ \citenamefont
  {Wang}}]{Wang:2024yag}%
  \BibitemOpen
  \bibfield  {author} {\bibinfo {author} {\bibfnamefont {S.}~\bibnamefont
  {Wang}}, \bibinfo {author} {\bibfnamefont {S.}~\bibnamefont {Li}}, \bibinfo
  {author} {\bibfnamefont {Y.}~\bibnamefont {Li}}, \bibinfo {author}
  {\bibfnamefont {B.-W.}\ \bibnamefont {Zhang}}, \ and\ \bibinfo {author}
  {\bibfnamefont {E.}~\bibnamefont {Wang}},\ }\href {\doibase
  10.1088/1674-1137/adb385} {\bibfield  {journal} {\bibinfo  {journal} {Chin.
  Phys. C}\ }\textbf {\bibinfo {volume} {49}},\ \bibinfo {pages} {064101}
  (\bibinfo {year} {2025}{\natexlab{a}})},\ \Eprint
  {http://arxiv.org/abs/2410.21834} {arXiv:2410.21834 [hep-ph]} \BibitemShut
  {NoStop}%
\bibitem [{\citenamefont {Li}\ \emph {et~al.}(2025)\citenamefont {Li},
  \citenamefont {Chen}, \citenamefont {Kong}, \citenamefont {Wang},\ and\
  \citenamefont {Zhang}}]{Li:2024pfi}%
  \BibitemOpen
  \bibfield  {author} {\bibinfo {author} {\bibfnamefont {Y.}~\bibnamefont
  {Li}}, \bibinfo {author} {\bibfnamefont {S.-Y.}\ \bibnamefont {Chen}},
  \bibinfo {author} {\bibfnamefont {W.-X.}\ \bibnamefont {Kong}}, \bibinfo
  {author} {\bibfnamefont {S.}~\bibnamefont {Wang}}, \ and\ \bibinfo {author}
  {\bibfnamefont {B.-W.}\ \bibnamefont {Zhang}},\ }\href {\doibase
  10.1088/0256-307X/42/1/011201} {\bibfield  {journal} {\bibinfo  {journal}
  {Chin. Phys. Lett.}\ }\textbf {\bibinfo {volume} {42}},\ \bibinfo {pages}
  {011201} (\bibinfo {year} {2025})},\ \Eprint
  {http://arxiv.org/abs/2409.12742} {arXiv:2409.12742 [hep-ph]} \BibitemShut
  {NoStop}%
\bibitem [{\citenamefont {Wang}\ \emph
  {et~al.}(2025{\natexlab{b}})\citenamefont {Wang}, \citenamefont {Li},
  \citenamefont {Shen}, \citenamefont {Zhang},\ and\ \citenamefont
  {Wang}}]{Wang:2023udp}%
  \BibitemOpen
  \bibfield  {author} {\bibinfo {author} {\bibfnamefont {S.}~\bibnamefont
  {Wang}}, \bibinfo {author} {\bibfnamefont {Y.}~\bibnamefont {Li}}, \bibinfo
  {author} {\bibfnamefont {S.}~\bibnamefont {Shen}}, \bibinfo {author}
  {\bibfnamefont {B.-W.}\ \bibnamefont {Zhang}}, \ and\ \bibinfo {author}
  {\bibfnamefont {E.}~\bibnamefont {Wang}},\ }\href {\doibase
  10.1103/PhysRevC.111.034912} {\bibfield  {journal} {\bibinfo  {journal}
  {Phys. Rev. C}\ }\textbf {\bibinfo {volume} {111}},\ \bibinfo {pages}
  {034912} (\bibinfo {year} {2025}{\natexlab{b}})},\ \Eprint
  {http://arxiv.org/abs/2308.14538} {arXiv:2308.14538 [hep-ph]} \BibitemShut
  {NoStop}%
\bibitem [{\citenamefont {Guo}\ and\ \citenamefont {Wang}(2000)}]{Guo:2000nz}%
  \BibitemOpen
  \bibfield  {author} {\bibinfo {author} {\bibfnamefont {X.-f.}\ \bibnamefont
  {Guo}}\ and\ \bibinfo {author} {\bibfnamefont {X.-N.}\ \bibnamefont {Wang}},\
  }\href {\doibase 10.1103/PhysRevLett.85.3591} {\bibfield  {journal} {\bibinfo
   {journal} {Phys. Rev. Lett.}\ }\textbf {\bibinfo {volume} {85}},\ \bibinfo
  {pages} {3591} (\bibinfo {year} {2000})},\ \Eprint
  {http://arxiv.org/abs/hep-ph/0005044} {arXiv:hep-ph/0005044} \BibitemShut
  {NoStop}%
\bibitem [{\citenamefont {Zhang}\ and\ \citenamefont
  {Wang}(2003)}]{Zhang:2003yn}%
  \BibitemOpen
  \bibfield  {author} {\bibinfo {author} {\bibfnamefont {B.-W.}\ \bibnamefont
  {Zhang}}\ and\ \bibinfo {author} {\bibfnamefont {X.-N.}\ \bibnamefont
  {Wang}},\ }\href {\doibase 10.1016/S0375-9474(03)01003-0} {\bibfield
  {journal} {\bibinfo  {journal} {Nucl. Phys. A}\ }\textbf {\bibinfo {volume}
  {720}},\ \bibinfo {pages} {429} (\bibinfo {year} {2003})},\ \Eprint
  {http://arxiv.org/abs/hep-ph/0301195} {arXiv:hep-ph/0301195} \BibitemShut
  {NoStop}%
\bibitem [{\citenamefont {Zhang}\ \emph {et~al.}(2004)\citenamefont {Zhang},
  \citenamefont {Wang},\ and\ \citenamefont {Wang}}]{Zhang:2003wk}%
  \BibitemOpen
  \bibfield  {author} {\bibinfo {author} {\bibfnamefont {B.-W.}\ \bibnamefont
  {Zhang}}, \bibinfo {author} {\bibfnamefont {E.}~\bibnamefont {Wang}}, \ and\
  \bibinfo {author} {\bibfnamefont {X.-N.}\ \bibnamefont {Wang}},\ }\href
  {\doibase 10.1103/PhysRevLett.93.072301} {\bibfield  {journal} {\bibinfo
  {journal} {Phys. Rev. Lett.}\ }\textbf {\bibinfo {volume} {93}},\ \bibinfo
  {pages} {072301} (\bibinfo {year} {2004})},\ \Eprint
  {http://arxiv.org/abs/nucl-th/0309040} {arXiv:nucl-th/0309040} \BibitemShut
  {NoStop}%
\bibitem [{\citenamefont {Majumder}(2012)}]{Majumder:2009ge}%
  \BibitemOpen
  \bibfield  {author} {\bibinfo {author} {\bibfnamefont {A.}~\bibnamefont
  {Majumder}},\ }\href {\doibase 10.1103/PhysRevD.85.014023} {\bibfield
  {journal} {\bibinfo  {journal} {Phys. Rev. D}\ }\textbf {\bibinfo {volume}
  {85}},\ \bibinfo {pages} {014023} (\bibinfo {year} {2012})},\ \Eprint
  {http://arxiv.org/abs/0912.2987} {arXiv:0912.2987 [nucl-th]} \BibitemShut
  {NoStop}%
\bibitem [{\citenamefont {Deng}\ and\ \citenamefont
  {Wang}(2010)}]{Deng:2009ncl}%
  \BibitemOpen
  \bibfield  {author} {\bibinfo {author} {\bibfnamefont {W.-t.}\ \bibnamefont
  {Deng}}\ and\ \bibinfo {author} {\bibfnamefont {X.-N.}\ \bibnamefont
  {Wang}},\ }\href {\doibase 10.1103/PhysRevC.81.024902} {\bibfield  {journal}
  {\bibinfo  {journal} {Phys. Rev. C}\ }\textbf {\bibinfo {volume} {81}},\
  \bibinfo {pages} {024902} (\bibinfo {year} {2010})},\ \Eprint
  {http://arxiv.org/abs/0910.3403} {arXiv:0910.3403 [hep-ph]} \BibitemShut
  {NoStop}%
\bibitem [{\citenamefont {He}\ \emph {et~al.}(2012)\citenamefont {He},
  \citenamefont {Fries},\ and\ \citenamefont {Rapp}}]{He:2011zx}%
  \BibitemOpen
  \bibfield  {author} {\bibinfo {author} {\bibfnamefont {M.}~\bibnamefont
  {He}}, \bibinfo {author} {\bibfnamefont {R.~J.}\ \bibnamefont {Fries}}, \
  and\ \bibinfo {author} {\bibfnamefont {R.}~\bibnamefont {Rapp}},\ }\href
  {\doibase 10.1103/PhysRevC.85.044911} {\bibfield  {journal} {\bibinfo
  {journal} {Phys. Rev. C}\ }\textbf {\bibinfo {volume} {85}},\ \bibinfo
  {pages} {044911} (\bibinfo {year} {2012})},\ \Eprint
  {http://arxiv.org/abs/1112.5894} {arXiv:1112.5894 [nucl-th]} \BibitemShut
  {NoStop}%
\bibitem [{\citenamefont {Wang}\ \emph {et~al.}(1995)\citenamefont {Wang},
  \citenamefont {Gyulassy},\ and\ \citenamefont {Plumer}}]{Wang:1994fx}%
  \BibitemOpen
  \bibfield  {author} {\bibinfo {author} {\bibfnamefont {X.-N.}\ \bibnamefont
  {Wang}}, \bibinfo {author} {\bibfnamefont {M.}~\bibnamefont {Gyulassy}}, \
  and\ \bibinfo {author} {\bibfnamefont {M.}~\bibnamefont {Plumer}},\ }\href
  {\doibase 10.1103/PhysRevD.51.3436} {\bibfield  {journal} {\bibinfo
  {journal} {Phys. Rev. D}\ }\textbf {\bibinfo {volume} {51}},\ \bibinfo
  {pages} {3436} (\bibinfo {year} {1995})},\ \Eprint
  {http://arxiv.org/abs/hep-ph/9408344} {arXiv:hep-ph/9408344} \BibitemShut
  {NoStop}%
\bibitem [{\citenamefont {Zakharov}(1996)}]{Zakharov:1996fv}%
  \BibitemOpen
  \bibfield  {author} {\bibinfo {author} {\bibfnamefont {B.~G.}\ \bibnamefont
  {Zakharov}},\ }\href {\doibase 10.1134/1.567126} {\bibfield  {journal}
  {\bibinfo  {journal} {JETP Lett.}\ }\textbf {\bibinfo {volume} {63}},\
  \bibinfo {pages} {952} (\bibinfo {year} {1996})},\ \Eprint
  {http://arxiv.org/abs/hep-ph/9607440} {arXiv:hep-ph/9607440} \BibitemShut
  {NoStop}%
\bibitem [{\citenamefont {Chen}\ \emph {et~al.}(2010)\citenamefont {Chen},
  \citenamefont {Greiner}, \citenamefont {Wang}, \citenamefont {Wang},\ and\
  \citenamefont {Xu}}]{Chen:2010te}%
  \BibitemOpen
  \bibfield  {author} {\bibinfo {author} {\bibfnamefont {X.-F.}\ \bibnamefont
  {Chen}}, \bibinfo {author} {\bibfnamefont {C.}~\bibnamefont {Greiner}},
  \bibinfo {author} {\bibfnamefont {E.}~\bibnamefont {Wang}}, \bibinfo {author}
  {\bibfnamefont {X.-N.}\ \bibnamefont {Wang}}, \ and\ \bibinfo {author}
  {\bibfnamefont {Z.}~\bibnamefont {Xu}},\ }\href {\doibase
  10.1103/PhysRevC.81.064908} {\bibfield  {journal} {\bibinfo  {journal} {Phys.
  Rev. C}\ }\textbf {\bibinfo {volume} {81}},\ \bibinfo {pages} {064908}
  (\bibinfo {year} {2010})},\ \Eprint {http://arxiv.org/abs/1002.1165}
  {arXiv:1002.1165 [nucl-th]} \BibitemShut {NoStop}%
\bibitem [{\citenamefont {Beraudo}\ \emph {et~al.}(2018)\citenamefont {Beraudo}
  \emph {et~al.}}]{Rapp:2018qla}%
  \BibitemOpen
  \bibfield  {author} {\bibinfo {author} {\bibfnamefont {A.}~\bibnamefont
  {Beraudo}} \emph {et~al.},\ }\href {\doibase 10.1016/j.nuclphysa.2018.09.002}
  {\bibfield  {journal} {\bibinfo  {journal} {Nucl. Phys. A}\ }\textbf
  {\bibinfo {volume} {979}},\ \bibinfo {pages} {21} (\bibinfo {year} {2018})},\
  \Eprint {http://arxiv.org/abs/1803.03824} {arXiv:1803.03824 [nucl-th]}
  \BibitemShut {NoStop}%
\bibitem [{\citenamefont {Cao}\ \emph {et~al.}(2019)\citenamefont {Cao} \emph
  {et~al.}}]{Cao:2018ews}%
  \BibitemOpen
  \bibfield  {author} {\bibinfo {author} {\bibfnamefont {S.}~\bibnamefont
  {Cao}} \emph {et~al.},\ }\href {\doibase 10.1103/PhysRevC.99.054907}
  {\bibfield  {journal} {\bibinfo  {journal} {Phys. Rev. C}\ }\textbf {\bibinfo
  {volume} {99}},\ \bibinfo {pages} {054907} (\bibinfo {year} {2019})},\
  \Eprint {http://arxiv.org/abs/1809.07894} {arXiv:1809.07894 [nucl-th]}
  \BibitemShut {NoStop}%
\bibitem [{\citenamefont {Ma}\ \emph {et~al.}(2019)\citenamefont {Ma},
  \citenamefont {Dai}, \citenamefont {Zhang},\ and\ \citenamefont
  {Wang}}]{Ma:2018swx}%
  \BibitemOpen
  \bibfield  {author} {\bibinfo {author} {\bibfnamefont {G.-Y.}\ \bibnamefont
  {Ma}}, \bibinfo {author} {\bibfnamefont {W.}~\bibnamefont {Dai}}, \bibinfo
  {author} {\bibfnamefont {B.-W.}\ \bibnamefont {Zhang}}, \ and\ \bibinfo
  {author} {\bibfnamefont {E.-K.}\ \bibnamefont {Wang}},\ }\href {\doibase
  10.1140/epjc/s10052-019-7005-z} {\bibfield  {journal} {\bibinfo  {journal}
  {Eur. Phys. J. C}\ }\textbf {\bibinfo {volume} {79}},\ \bibinfo {pages} {518}
  (\bibinfo {year} {2019})},\ \Eprint {http://arxiv.org/abs/1812.02033}
  {arXiv:1812.02033 [nucl-th]} \BibitemShut {NoStop}%
\bibitem [{\citenamefont {Braaten}\ and\ \citenamefont
  {Thoma}(1991)}]{Braaten:1991we}%
  \BibitemOpen
  \bibfield  {author} {\bibinfo {author} {\bibfnamefont {E.}~\bibnamefont
  {Braaten}}\ and\ \bibinfo {author} {\bibfnamefont {M.~H.}\ \bibnamefont
  {Thoma}},\ }\href {\doibase 10.1103/PhysRevD.44.R2625} {\bibfield  {journal}
  {\bibinfo  {journal} {Phys. Rev. D}\ }\textbf {\bibinfo {volume} {44}},\
  \bibinfo {pages} {R2625} (\bibinfo {year} {1991})}\BibitemShut {NoStop}%
\bibitem [{\citenamefont {Neufeld}(2011)}]{Neufeld:2010xi}%
  \BibitemOpen
  \bibfield  {author} {\bibinfo {author} {\bibfnamefont {R.~B.}\ \bibnamefont
  {Neufeld}},\ }\href {\doibase 10.1103/PhysRevD.83.065012} {\bibfield
  {journal} {\bibinfo  {journal} {Phys. Rev. D}\ }\textbf {\bibinfo {volume}
  {83}},\ \bibinfo {pages} {065012} (\bibinfo {year} {2011})},\ \Eprint
  {http://arxiv.org/abs/1011.4979} {arXiv:1011.4979 [hep-ph]} \BibitemShut
  {NoStop}%
\bibitem [{\citenamefont {Miller}\ \emph {et~al.}(2007)\citenamefont {Miller},
  \citenamefont {Reygers}, \citenamefont {Sanders},\ and\ \citenamefont
  {Steinberg}}]{Miller:2007ri}%
  \BibitemOpen
  \bibfield  {author} {\bibinfo {author} {\bibfnamefont {M.~L.}\ \bibnamefont
  {Miller}}, \bibinfo {author} {\bibfnamefont {K.}~\bibnamefont {Reygers}},
  \bibinfo {author} {\bibfnamefont {S.~J.}\ \bibnamefont {Sanders}}, \ and\
  \bibinfo {author} {\bibfnamefont {P.}~\bibnamefont {Steinberg}},\ }\href
  {\doibase 10.1146/annurev.nucl.57.090506.123020} {\bibfield  {journal}
  {\bibinfo  {journal} {Ann. Rev. Nucl. Part. Sci.}\ }\textbf {\bibinfo
  {volume} {57}},\ \bibinfo {pages} {205} (\bibinfo {year} {2007})},\ \Eprint
  {http://arxiv.org/abs/nucl-ex/0701025} {arXiv:nucl-ex/0701025} \BibitemShut
  {NoStop}%
\bibitem [{\citenamefont {Pang}\ \emph {et~al.}(2016)\citenamefont {Pang},
  \citenamefont {Petersen}, \citenamefont {Wang},\ and\ \citenamefont
  {Wang}}]{Pang:2016igs}%
  \BibitemOpen
  \bibfield  {author} {\bibinfo {author} {\bibfnamefont {L.-G.}\ \bibnamefont
  {Pang}}, \bibinfo {author} {\bibfnamefont {H.}~\bibnamefont {Petersen}},
  \bibinfo {author} {\bibfnamefont {Q.}~\bibnamefont {Wang}}, \ and\ \bibinfo
  {author} {\bibfnamefont {X.-N.}\ \bibnamefont {Wang}},\ }\href {\doibase
  10.1103/PhysRevLett.117.192301} {\bibfield  {journal} {\bibinfo  {journal}
  {Phys. Rev. Lett.}\ }\textbf {\bibinfo {volume} {117}},\ \bibinfo {pages}
  {192301} (\bibinfo {year} {2016})},\ \Eprint
  {http://arxiv.org/abs/1605.04024} {arXiv:1605.04024 [hep-ph]} \BibitemShut
  {NoStop}%
\bibitem [{\citenamefont {Putschke}\ \emph {et~al.}(2019)\citenamefont
  {Putschke} \emph {et~al.}}]{Putschke:2019yrg}%
  \BibitemOpen
  \bibfield  {author} {\bibinfo {author} {\bibfnamefont {J.~H.}\ \bibnamefont
  {Putschke}} \emph {et~al.},\ }\href@noop {} {\  (\bibinfo {year} {2019})},\
  \Eprint {http://arxiv.org/abs/1903.07706} {arXiv:1903.07706 [nucl-th]}
  \BibitemShut {NoStop}%
\bibitem [{\citenamefont {Andersson}\ \emph {et~al.}(1983)\citenamefont
  {Andersson}, \citenamefont {Gustafson},\ and\ \citenamefont
  {Soderberg}}]{Andersson:1983jt}%
  \BibitemOpen
  \bibfield  {author} {\bibinfo {author} {\bibfnamefont {B.}~\bibnamefont
  {Andersson}}, \bibinfo {author} {\bibfnamefont {G.}~\bibnamefont
  {Gustafson}}, \ and\ \bibinfo {author} {\bibfnamefont {B.}~\bibnamefont
  {Soderberg}},\ }\href {\doibase 10.1007/BF01407824} {\bibfield  {journal}
  {\bibinfo  {journal} {Z. Phys. C}\ }\textbf {\bibinfo {volume} {20}},\
  \bibinfo {pages} {317} (\bibinfo {year} {1983})}\BibitemShut {NoStop}%
\bibitem [{\citenamefont {Sjostrand}(1984)}]{Sjostrand:1984ic}%
  \BibitemOpen
  \bibfield  {author} {\bibinfo {author} {\bibfnamefont {T.}~\bibnamefont
  {Sjostrand}},\ }\href {\doibase 10.1016/0550-3213(84)90607-2} {\bibfield
  {journal} {\bibinfo  {journal} {Nucl. Phys. B}\ }\textbf {\bibinfo {volume}
  {248}},\ \bibinfo {pages} {469} (\bibinfo {year} {1984})}\BibitemShut
  {NoStop}%
\bibitem [{\citenamefont {Cooper}\ and\ \citenamefont
  {Frye}(1974)}]{Cooper:1974mv}%
  \BibitemOpen
  \bibfield  {author} {\bibinfo {author} {\bibfnamefont {F.}~\bibnamefont
  {Cooper}}\ and\ \bibinfo {author} {\bibfnamefont {G.}~\bibnamefont {Frye}},\
  }\href {\doibase 10.1103/PhysRevD.10.186} {\bibfield  {journal} {\bibinfo
  {journal} {Phys. Rev. D}\ }\textbf {\bibinfo {volume} {10}},\ \bibinfo
  {pages} {186} (\bibinfo {year} {1974})}\BibitemShut {NoStop}%
\bibitem [{\citenamefont {Abelev}\ \emph {et~al.}(2013)\citenamefont {Abelev}
  \emph {et~al.}}]{ALICE:2013mez}%
  \BibitemOpen
  \bibfield  {author} {\bibinfo {author} {\bibfnamefont {B.}~\bibnamefont
  {Abelev}} \emph {et~al.} (\bibinfo {collaboration} {ALICE}),\ }\href
  {\doibase 10.1103/PhysRevC.88.044910} {\bibfield  {journal} {\bibinfo
  {journal} {Phys. Rev. C}\ }\textbf {\bibinfo {volume} {88}},\ \bibinfo
  {pages} {044910} (\bibinfo {year} {2013})},\ \Eprint
  {http://arxiv.org/abs/1303.0737} {arXiv:1303.0737 [hep-ex]} \BibitemShut
  {NoStop}%
\bibitem [{\citenamefont {Cacciari}\ \emph {et~al.}(2008)\citenamefont
  {Cacciari}, \citenamefont {Salam},\ and\ \citenamefont
  {Soyez}}]{Cacciari:2008gp}%
  \BibitemOpen
  \bibfield  {author} {\bibinfo {author} {\bibfnamefont {M.}~\bibnamefont
  {Cacciari}}, \bibinfo {author} {\bibfnamefont {G.~P.}\ \bibnamefont {Salam}},
  \ and\ \bibinfo {author} {\bibfnamefont {G.}~\bibnamefont {Soyez}},\ }\href
  {\doibase 10.1088/1126-6708/2008/04/063} {\bibfield  {journal} {\bibinfo
  {journal} {JHEP}\ }\textbf {\bibinfo {volume} {04}},\ \bibinfo {pages} {063}
  (\bibinfo {year} {2008})},\ \Eprint {http://arxiv.org/abs/0802.1189}
  {arXiv:0802.1189 [hep-ph]} \BibitemShut {NoStop}%
\bibitem [{\citenamefont {Yan}\ \emph {et~al.}(2021)\citenamefont {Yan},
  \citenamefont {Chen}, \citenamefont {Dai}, \citenamefont {Zhang},\ and\
  \citenamefont {Wang}}]{Yan:2020zrz}%
  \BibitemOpen
  \bibfield  {author} {\bibinfo {author} {\bibfnamefont {J.}~\bibnamefont
  {Yan}}, \bibinfo {author} {\bibfnamefont {S.-Y.}\ \bibnamefont {Chen}},
  \bibinfo {author} {\bibfnamefont {W.}~\bibnamefont {Dai}}, \bibinfo {author}
  {\bibfnamefont {B.-W.}\ \bibnamefont {Zhang}}, \ and\ \bibinfo {author}
  {\bibfnamefont {E.}~\bibnamefont {Wang}},\ }\href {\doibase
  10.1088/1674-1137/abca2b} {\bibfield  {journal} {\bibinfo  {journal} {Chin.
  Phys. C}\ }\textbf {\bibinfo {volume} {45}},\ \bibinfo {pages} {024102}
  (\bibinfo {year} {2021})},\ \Eprint {http://arxiv.org/abs/2005.01093}
  {arXiv:2005.01093 [hep-ph]} \BibitemShut {NoStop}%
\bibitem [{\citenamefont {Cal}\ \emph {et~al.}(2020)\citenamefont {Cal},
  \citenamefont {Neill}, \citenamefont {Ringer},\ and\ \citenamefont
  {Waalewijn}}]{Cal:2019gxa}%
  \BibitemOpen
  \bibfield  {author} {\bibinfo {author} {\bibfnamefont {P.}~\bibnamefont
  {Cal}}, \bibinfo {author} {\bibfnamefont {D.}~\bibnamefont {Neill}}, \bibinfo
  {author} {\bibfnamefont {F.}~\bibnamefont {Ringer}}, \ and\ \bibinfo {author}
  {\bibfnamefont {W.~J.}\ \bibnamefont {Waalewijn}},\ }\href {\doibase
  10.1007/JHEP04(2020)211} {\bibfield  {journal} {\bibinfo  {journal} {JHEP}\
  }\textbf {\bibinfo {volume} {04}},\ \bibinfo {pages} {211} (\bibinfo {year}
  {2020})},\ \Eprint {http://arxiv.org/abs/1911.06840} {arXiv:1911.06840
  [hep-ph]} \BibitemShut {NoStop}%
\bibitem [{\citenamefont {Bertolini}\ \emph {et~al.}(2014)\citenamefont
  {Bertolini}, \citenamefont {Chan},\ and\ \citenamefont
  {Thaler}}]{Bertolini:2013iqa}%
  \BibitemOpen
  \bibfield  {author} {\bibinfo {author} {\bibfnamefont {D.}~\bibnamefont
  {Bertolini}}, \bibinfo {author} {\bibfnamefont {T.}~\bibnamefont {Chan}}, \
  and\ \bibinfo {author} {\bibfnamefont {J.}~\bibnamefont {Thaler}},\ }\href
  {\doibase 10.1007/JHEP04(2014)013} {\bibfield  {journal} {\bibinfo  {journal}
  {JHEP}\ }\textbf {\bibinfo {volume} {04}},\ \bibinfo {pages} {013} (\bibinfo
  {year} {2014})},\ \Eprint {http://arxiv.org/abs/1310.7584} {arXiv:1310.7584
  [hep-ph]} \BibitemShut {NoStop}%
\bibitem [{\citenamefont {Acharya}\ \emph
  {et~al.}(2024{\natexlab{a}})\citenamefont {Acharya} \emph
  {et~al.}}]{ALICE:2023qve}%
  \BibitemOpen
  \bibfield  {author} {\bibinfo {author} {\bibfnamefont {S.}~\bibnamefont
  {Acharya}} \emph {et~al.} (\bibinfo {collaboration} {ALICE}),\ }\href
  {\doibase 10.1103/PhysRevLett.133.022301} {\bibfield  {journal} {\bibinfo
  {journal} {Phys. Rev. Lett.}\ }\textbf {\bibinfo {volume} {133}},\ \bibinfo
  {pages} {022301} (\bibinfo {year} {2024}{\natexlab{a}})},\ \Eprint
  {http://arxiv.org/abs/2308.16131} {arXiv:2308.16131 [nucl-ex]} \BibitemShut
  {NoStop}%
\bibitem [{\citenamefont {Acharya}\ \emph
  {et~al.}(2024{\natexlab{b}})\citenamefont {Acharya} \emph
  {et~al.}}]{ALICE:2023jye}%
  \BibitemOpen
  \bibfield  {author} {\bibinfo {author} {\bibfnamefont {S.}~\bibnamefont
  {Acharya}} \emph {et~al.} (\bibinfo {collaboration} {ALICE}),\ }\href
  {\doibase 10.1103/PhysRevC.110.014906} {\bibfield  {journal} {\bibinfo
  {journal} {Phys. Rev. C}\ }\textbf {\bibinfo {volume} {110}},\ \bibinfo
  {pages} {014906} (\bibinfo {year} {2024}{\natexlab{b}})},\ \Eprint
  {http://arxiv.org/abs/2308.16128} {arXiv:2308.16128 [nucl-ex]} \BibitemShut
  {NoStop}%
\end{thebibliography}%

\end{document}